\documentclass{aastex631}
\submitjournal{ApJ (Accepted 5 July)}

\shorttitle{The $\Gamma(a,bt)$ CSFRD evolution of the Universe.}
\shortauthors{Katsianis et al.}

\begin{document}

\title{The observed cosmic star formation rate density has an evolution which resembles a $\Gamma(a,bt)$ distribution and can be described successfully  by only 2 parameters.}

\correspondingauthor{Antonios Katsianis}
\email{kata@sjtu.edu.cn}

\author{Antonios Katsianis}
\affiliation{Tsung-Dao Lee Institute, Shanghai Jiao Tong University, Shanghai 200240, China}
\affiliation{Department of Astronomy, Shanghai Key Laboratory for Particle Physics and Cosmology, Shanghai Jiao Tong University, Shanghai 200240, China}

\author{Xiaohu Yang}
\affiliation{Tsung-Dao Lee Institute, Shanghai Jiao Tong University, Shanghai 200240, China}
\affiliation{Department of Astronomy, Shanghai Key Laboratory for Particle Physics and Cosmology, Shanghai Jiao Tong University, Shanghai 200240, China}

\author{Xianzhong Zheng}
\affiliation{Purple Mountain Observatory, No. 10 Yuanhua Road, Nanjing 210023, China}

\begin{abstract}

  A debate is emerging regarding the recent inconsistent results of different studies for the Cosmic Star Formation Rate Density (CSFRD) at high-z. We employ UV and IR datasets to investigate the star formation rate function (SFRF) at ${\rm z \sim  0-9}$. We find that the SFRFs derived  from the dust corrected ${\rm UV}$ (${\rm UV_{corr}}$) data contradict those from IR on some key issues since they are described by different distributions (Schechter vs double-power law), imply different physics for galaxy formation (${\rm UV_{corr}}$ data suggest a SFR limit/strong mechanism that diminish the number density of high star forming systems with respect IR) and compare differently with the stellar mass density evolution obtained from SED fitting (${\rm UV_{corr}}$ is in agreement, while IR in tension up to 0.5 dex). However, both tracers agree on a constant CSFRD evolution at ${\rm z \sim 1-4}$ and point to a plateau instead of a peak. In addition, using both indicators we demonstrate that the evolution of the {\it observed} CSFRD can be described by only {\bf 2} parameters and a function that has the form of a Gamma distribution (${\bf \Gamma(a,bt)}$). In contrast to previous parameterizations used in the literature our framework connects the parameters to physical properties like the star formation rate depletion time and cosmic baryonic gas density. The build up of stellar mass occurs in $\Gamma(a,bt)$ distributed steps and is the result of gas consumption up to the limit that there is no eligible gas for SF at t = ${\rm \infty}$, resulting to a final cosmic stellar mass density of $\sim 0.5 \times 10^9 \, {\rm \frac{M_{\odot}}{Mpc^3}}$.

\end{abstract}


\section{Introduction}
\label{intro}

In the last 3 decades galaxy surveys and cosmological simulations have been used in our quest to understand how galaxies evolve. A particular focus has been given to the observed SFRs and stellar masses of galaxies which enable us to establish numerous constraints like the galaxy stellar mass function, the star formation rate function (SFRF), the star formation rate-stellar mass relation (${\rm SFR-M_{\star}}$), the cosmic star formation rate density (CSFRD) and the cosmic stellar mass density \citep{Lapo2017,Fernandez2018,Driver2018,Davies2019,Caplar2019,Blanc2019,Davies2019,Katsianis2019,Hodge2020,Cheng2020,Trcka2020,Tacchella2020,Thorne2020,Lovell2021,Vijayan2021}. 

However, different observational studies have been employing different methodologies/wavelengths in order to derive galaxy SFRs like IR luminosities \citep{Guo2015,Qin2019}, H$\alpha$ luminosities \citep{Sanchez2018,Cano2019}, the Spectral Energy fitting (SED) technique \citep{Kurczynski2017,Zhaoka2020,Yang2021} or UV luminosities \citep{Blanc2019,Moutard2020}. All the above methods have been commonly used in the literature and they have provided a great opportunity to study galaxy evolution but they suffer from various shortcomings. For example, UV is a  direct measurement of SFR but is limited as follows:

\begin{itemize}
\item Dust corrections for UV luminosities are uncertain for high star-forming systems and possibly underestimated \citep{Dunlop2017}, 
\item UV-LFs and conclusively the UV-SFRFs are probably incomplete at the bright end of the distribution since  bright objects/high star forming galaxies have high dust contents and thus may be invisible to UV surveys \citep{Katsianis2016},
\item  It is currently challenging to confirm how many stars a galaxy forms since the individual stellar populations cannot be resolved \citep{Kuncarayakti2016}. We do not have a cosmic timer to mark the time for the birth of stars and a cosmic scale to weight their masses so we cannot know if the actual star formation rates we derived from their UV luminosities are actually correct. Any calibration for the UV-SFR conversion derived from stellar population synthesis modeling (SPS) like those from \citet{kennicutt1998} or \citet{Kennicutt2012} relies on modeling and assumptions \citep{Chen2010,Stanway2020} \footnote{We note that individual stellar populations can be resolved for nearby objects. \citet{Olsen2021} using a sample of 36 nearby dwarf galaxies demonstrated a broad agreement between the star formation histories derived from the SEDs and the color magnitude diagrams.}. 
\end{itemize}

On the other hand, IR studies are traditionally considered to be able to probe the SFRs of high star forming systems. We have to note the following shortcomings of deriving ${\rm SFR}$ and conclusively CSFRDs from IR data \citep[e.g.][]{Fumagalli2014,Utomo2014,Hayward2014,Katsianis2015,Katsianis2020} which can be summarized as:

\begin{itemize}  
\item  Overestimation due to buried AGNs that boosts the IR luminosities \citep{Brand2006,Ichikawa2012,Roebuck2016,Brown2019,Symeonidis2021},
\item overestimation due to the fact that dust can be heated by old populations not relevant  to current star-formation \citep{Viaene2017,Nersesian2019,Brown2019,Leja2019}. The IR luminosity can overestimate the instantaneous SFR during the post-starburst phase since stars that were formed in the starburst phase can remain dust-obscured and thus produce a significant IR luminosity non related to new born stars \citep{Hayward2014},
\item overestimation due to larger polycyclic aromatic hydrocarbons emission of distant galaxies \citep{Huang2009,Murata2014},
\item great uncertainties for the number density of galaxies with low SFRs (defined in our work as $SFR =  0.01-0.5$ \, ${\rm M_{\odot}/yr}$) and intermediate SFRs (defined as $SFR = 0.5-10$ ${\rm M_{\odot}/yr}$) which IR data cannot usually probe. This limitation is redshift dependent and deriving any parameterization of the total SFRF or total Luminosity function using only IR data (i.e. the high star forming end defined as objects with $SFR$ $>$ 10  ${\rm M_{\odot}/yr}$) can be problematic \citep{Katsianis2016,Katsianis2017},
\item high star forming/IR bright systems might be contaminated by the effect of gravitational lensing at high redshifts \citet{Zavala2021}, 
\item  insufficient wavelength coverage especially at FIR wavelengths \citep{Pearson2018} 
\item ultra-luminous IR galaxies are offset from the typically used SFR calibrations \citep{DeLooze2014},
\item IR estimations represent less instantaneous measurements than UV or H$\alpha$ tracers. Both UV and H$\alpha$ luminosities are sensitive only to the most massive stars which have short lifetimes \citep{Shivaei2015,Katsianis2015},
\item Like for the case of the UV SFRs it is currently challenging to confirm how many stars a high redshift galaxy forms per year. We stress that IR data are always important in order to build better SED models since they span an interesting part of the spectrum related to dust physics.
\end{itemize}

A natural question arises: 1) Since different indicators and methods rely on different principles (e.g. UV calibrations directly trace photons relevant to SF, while IR focus on the reprocessed light from dust) and have different shortcomings (e.g. UV light is affected by dust, while IR is contaminated by older population of stars) are they consistent with each other ?

In the last 5 years an increasing number of authors have reported a severe discrepancy between the SFRs inferred by different methodologies \citep{Hayward2014,Katsianis2014,Davies2016,Martis2019,Katsianis2020} in contrast with others that insist on a consistent picture of galaxy SFRs among different studies and indicators \citep{Dominguez2012,Madau2014,Rodighiero2014,Santini2017}. A conclusive study is required to establish if there is indeed a tension (and if yes estimate its magnitude). We have to keep in mind that numerous efforts have been done in the last 2 years that point shortcomings of previous published work/methodologies that rely on the traditionally used UV, IR or SED SFR indicators. Nowadays, there is evidence that previous stellar masses reported in the literature may have been underestimated by ${\rm 0.1 - 0.3}$ dex, while previous calculated SFRs may have been overestimated by ${\rm \sim 0.1}$ - ${\rm 1.0}$ dex \citep{Leja2019b,Katsianis2020}. 

Despite the above limitations of the traditionally used IR and UV indicators, recently a range of IR/CO studies \citep{Rowanrobinson2016,Loiacono2020,Khusanova2020,Gruppioni2020} suggest that a considerable amount of star forming activity occurs at high-z that is not recovered by UV data {\it while} cosmological simulations like EAGLE \citep{Schaye2015,Lagos2020} or IllustrisTNG \citep{Pillepich2018,KopenhaferClaire2020} underpredict the CSFRD with respect their observations. Thus, models and previously reported UV, H$\alpha$ and SED Cosmic SFRs are challenged by the authors. However, we present in the previous paragraphs that {\it all} indicators and methods  possibly suffer from limitations and thus the above statement can be reversed by some authors to: The IR/CO studies mentioned above involve CSFRDs that are overestimated and cosmological models, simulations, UV, SED and H$\alpha$ measurements are representing better the ``true'' SFRs at high-z.

A  powerful  way  to investigate the ability of an observational tool in deriving physical quantities is by utilizing  dust  radiative  transfer models and simulations \citep{Dickey2020,Baes2020,LuYi2020,Narayan2021,Lovell2021}. Recent studies investigated morphology  measures \citep{Cochrane2019}, star formation rate indicators \citep{Katsianis2020,Lower2020} or galaxy dust attenuation curves \citep{Trayford2020}. More specifically, \citet{Katsianis2020} suggested that for high redshifts UV/SED/H$\alpha$ SFR measurements are relatively robust, while numerous calibrations that rely on a combination of UV and IR luminosities \citep{Heinis2014,Whitaker2014} overproduce the derived SFRs even by 0.5 dex, resulting in an illusionary tension between observed and simulated ${\rm SFR-M_{\star}}$ relations. However, we stress that we cannot claim conclusively if the one or the other technique is better for real galaxies, since cosmological models suffer as well from numerous shortcomings. For example, the SFRs and stellar masses obtained by state-of-the-art cosmological simulations like EAGLE or IllustrisTNG and any post-processing with radiative transfer codes can be affected by:
\begin{itemize}
\item Resolution effects. It has been demonstrated that the simulated SFRFs and stellar mass functions are not converging among runs of different resolution \citep{Schaye2015,Pillepich2018,Zhaoka2020}. In order for convergence to be achieved a resolution dependent re-tuning of the sub-grid prescriptions (e.g. Feedback\footnote{We note that there are great uncertainties for the simulated sSFRs (the $SFR/M_{\star}$ ratio) of passive galaxies ($sSFR < 10^{-11} \, M_{\odot}/yr$). The number density of the specific star formation rates, namely the specific star formation rate function predicted from the simulations is in severe tension with observations \citep{Zhaoka2020,Katsianis2021,CorchoCab2021}. This disagreement hints that galaxies are not quenched correctly in the simulations via Feedback and the results from the models should be treated with caution.}) is required and this rises the question if our models reproduce galaxy properties for the right reasons. In addition, resolution limits cause uncertainties for the post-processing of the simulated galaxies via radiative transfer \citep{Trayford2017,Nelsoncolor}. For example, the luminosity at mid-infrared wavelengths mainly originates from star forming regions which are well below the resolution limits of the current state-of-the-art simulations so further uncertain sub-grid modeling is required \citep{Baes2020}.
\item Any comparison in the observed flux space (mock observations that are built upon the cosmological simulations) is always affected by the assumptions employed by the radiative transfer post-processing e.g. the assumed dust model \citep{Calzetti1994,zubko2004,Nelsoncolor} or the adopted metal fraction \citep{Brinchmann2013,Camps2018}. 
\end{itemize}

Besides these limitations though it is important to note that simulations/radiative transfer have progressed significantly and can provide strong hints and an insight of the shortcomings of tools that are being used to derive physical quantities from observations. Keeping all in mind, the question that arises this time is: 2) Besides using simulations/radiative transfer is there a way that we can determine which method describes more successfully  the ``true'' SFRs ?

Severe limitations exist in the field of theory of star formation as well and especially in the parameterization of the star formation histories of both individual galaxies and of the CSFRD. Besides the fact that the SFHs of realistic  galaxies  can  be complex, they are often modeled with simple functional forms.  These forms, besides their limitations \citep{SmithDaniel2015,Carnall2019,Leja2019}, provide a computationally fast approach and are able to be applied to SED fitting procedures \citep{Leja2017}.  For the case of individual galaxies, the most common scenario is an exponential form with $SFR \propto t^B \times e^{-t/\tau}$. When $B = 0$ the law describes a simple exponentially declining SFH and when $B = 1$ a delayed-exponentially  declining  SFH. More complex forms have been adopted to better describe the range of observations.  For example the rising \citep{Papovich2011}, log-normal \citep{Abramson2016} and  double  power-law  \citep{Behroozi2013} SFHs have been proposed. For the case of the cosmic star formation rate density the most commonly used parameterizations are given by \citet{Madau2014} who suggested that the cosmic SFH follows a rising phase, scaling as $SFR(z) \propto (1 + z)^{-2.9}$ at ${\rm z = 3 - 8}$, slowing down and peaking at ${\rm z \sim 2}$, followed by a gradual decline to the present day, roughly as ${\rm SFR(z) \propto (1 +z)^{2.7}}$. Similar models have been suggested earlier by \citet{Cole2001} as $CSFRD = \frac{a+b \, z}{1+(\frac{z}{c})^d}$. Two questions arise: 1) These parameterizations are empirically motivated and represent ``physics free'' models. Can the parameters be connected to the physical properties of galaxies ? 2) Is it possible to decrease the number of parameters necessary (4) to broadly fit the data ?

In our work, we construct star formation rate functions to demonstrate how different {\it qualitative} results the two schools of thought (IR vs UV) produce, pointing out the importance of the tension (subsection \ref{SFRF}). We perform sanity checks by making comparisons with other observables (the evolution of the cosmic stellar mass density at ${\rm z \sim 0-9}$) as we deem this process a complementary approach to radiative transfer/simulations to start uncovering which indicator is more consistent with the current paradigm of galaxy formation (section \ref{CSFRD}). Finally, we establish a strong simple paramerization that describes the CSFRD derived from the UV dust corrected (UV$_{Corr}$) and IR data using {\it only 2} parameters and a function that resembles a {\it Gamma distribution}. This parameterization connects its parameters to properties of galaxies/halos/Universe and thus it is an effort to extend the current ``physics free'' empirical fits (section \ref{Justification}). In our work we adopt a \citet{Planck2018} cosmology with  $\Omega_m = 0.315$, $\Omega_{\Lambda} = 0.685$, $n_s=0.965$, $h = 0.674$  and a \citet{chabrier03} IMF.

\section{The ``observed'' star formation rates of galaxies from UV and IR light}
\label{SFRS}

\subsection{The data}
\label{data}

Bellow we summarize the datasets used for this work and the methodology for deriving galaxy star formation rates from UV and IR light. We start with the observations of \citet{Moutard2020} who obtained the rest-frame FUV (1546 ${\rm \AA}$) luminosity functions at redshifts ${\rm z \sim 0.2}$ to ${\rm z \sim 3}$.  Their work spans over 4.3 million galaxies, selected  from  the Canada-France-Hawaii Telescope (CFHT)  Large  Area U-band  Deep  Survey  \citep[CLAUDS, ][]{Sawicki2019}  and  the HyperSuprime-Cam Subaru Strategic Program \citep[HSC-SSP, ][]{Aihara2018}. This combination of area and depth enabled the authors to probe both the faint end of the UV LF regime within excellent statistical precision, while also taking into account very rare galaxies at the bright end which are essentially free of cosmic variance. Thus, this work can provide an excellent starting point for determining the UV SFRFs and provide comparisons with the predictions from IR data that actually are supposed to probe mostly the most star forming/dusty objects at high-z. The authors noted that the LFs are described by classic Schechter forms. Besides \citet{Moutard2020} we employ additional UV LFs to extend our work to higher redshifts \citep{Ono2018,Bhatawdekar2019}. \citet{Ono2018} presented results for very luminous galaxies at ${\rm z \sim 4-7}$ based on the wide and deep optical Hyper Suprime-Cam while \citet{Bhatawdekar2019} gives insight for ${\rm z \sim 9-10}$ objects. Last we use the star formation rate functions presented in \citet{Katsianis2016} and \citet{Katsianis2017} that span a redshift range of ${\rm z \sim 0-8}$. The authors employed IR \citep{Patel13,Magnelli2013}, UV \citep{Bouwens2014,Parsa2015}, H$\alpha$ \citep{Sobral2013} and radio data \citep{Mauch2007}. We note that most of our knowledge for the cosmic star formation rate density (CSFRD) at high redshifts (z $>$ 2) is based mostly on galaxy samples selected in the rest frame UV \citep{Oesch2018}, while the total star formation rates (SFRs) are not measured, but rather inferred through dust-correction techniques. Following the commonly used IRX-${\rm \beta}$ relation \citep{meurer1999} as in \citet{smit12} and \citet{Katsianis2016,Katsianis2017} we correct the UV luminosities as follows:

\begin{eqnarray}
\label{eq_A16} 
{\rm A_{1600} = 4.43 + 1.99 \beta },
\end{eqnarray}
where $A_{1600}$ is the dust absorption at 1600 ${\rm \AA}$ and $\beta$ is the UV-continuum spectral slope. We assume a relation between $\beta$ and the luminosity \citep{bouwens2012,Tacchella2013}:

\begin{eqnarray}
  \label{eq_beta}
{\rm  \langle \beta \rangle = \frac{{\rm d}\beta}{{\rm d}M_{\rm UV}}\left(M_{\rm UV,AB}+19.5\right)+\beta_{M_{\rm UV}}},
\end{eqnarray}
we assume the same $\left < \beta \right >$ as \citet{Arnouts2005,Oesch2010,smit12,Tacchella2013,Katsianis2016} and \citet{Katsianis2020}. Then, following \citet{HaoKen} and \citet{Katsianis2017} we employ:

\begin{eqnarray}
\label{eq_A17}
{\rm L_{UV_{uncor}} = L_{UV_{corr}}e^{-\tau_{UV}}},
\end{eqnarray}

where $\tau_{UV}$ is the effective optical depth ($\tau_{UV} = A_{\rm 1600}/1.086$)$,  {\rm L_{UV_{uncor}}}$ is the observed (affected by dust luminosity) and ${\rm L_{UV_{corr}}}$ represents the intrinsic luminosity. We convert the dust-corrected UV luminosities into SFRs following \citet{Kennicutt2012}:

\begin{eqnarray}
\label{eq_A19}
{\rm Log_{10} (SFR_{UV,corr}) =  Log_{10} (L_{UV_{corr}}) - 43.35}.
\end{eqnarray}

We label the above determination as SFR${\rm_{UV,corr}}$. We note that besides the limitations of the indicator, its validity has been explored in cosmological simulations/radiative transfer codes in \citet{Katsianis2020} where it is shown that the ${\rm SFR_{UV, corr}}$ is successful at deriving any SFRs at ${\rm z \geq 2}$, but starts under performing at lower redshifts (${\rm z \le 1}$) for high star forming systems (${\rm > 10 \, M_{\odot} \, yr^{-1}}$)\footnote {We note that due to the shortcomings of both cosmological simulations and observational techniques, outlined in the introduction, it is uncertain if the SFR${\rm_{UV,corr}}$ indicator described above is not successful at ${\rm z \le 1}$ and ${SFR \rm > 10 \, M_{\odot} \, yr^{-1}}$ due to the fact that the dust is not modeled correctly in EAGLE or that indeed the method underestimates the SFRs of dusty galaxies at low redshifts.}. This confirmation via radiative transfer/simulations (besides their limitations) is encouraging for high-z/low SFR determinations of SFRs from UV light. We note that the above method produces typically results that are in good agreement with SFRs obtained from SED fitting and H$\alpha$ data.

In order to have a more complete picture of star formation it is a good practice to combine the UV SFRs with IR data. In our work for high redshift galaxies we employ the state-of-the-art data of \citet{Gruppioni2020}  who used 56 sources blindly detected within the ALPINE survey to investigate the evolution of the dusty high SFR galaxy population at ${\rm z \sim 0.5-6}$. The authors computed the rest-frame LFs at 250 ${\rm \mu m}$ and compared them with the Herschel and SCUBA-2 LFs suggesting that the ALPINE results are mostly complementary to the previous data. The authors computed the total IR luminosity by integrating the SEDs over 8-1000 ${\rm \mu m}$, constructed IR luminosity functions and employed the \citet{kennicutt1998} relation to obtain the evolution of the CSFRD. The results were found in agreement with those from previous far-IR data, CO LFs \citep{Riechers2019,Decarli2019} and GRB data \citep{Kistler2009}. \citet{Gruppioni2020} report that the CSFRD from their IR data are significantly higher than those found by optical/UV surveys at ${\rm z>2}$ even by a factor of about 10 at z = 6 claiming they are witnessing obscured star formation. Besides the small number of objects (56) the authors support that the area covered by the survey guarantees that the contribution due to cosmic variance to the derived LFs and CSFRD is negligible. However, according to \citet{Zavala2021} the CSFRD derived from \citet{Gruppioni2020} may be actually artificially overestimated due to numerous reasons as follows : 1) the fact that Herschel observations overestimate  the  derived  IR luminosities which conclusively overestimate the derived SFRs. 2) clustering effects which have their roots to the fact that the original targets are a few of massive galaxies at ${\rm z \sim 4-6}$ which represent an over-dense region of the Universe. 3) Uncertain extrapolations up to the faint end.

In addition to \citet{Gruppioni2020} in our study we employ the IR LFs from other studies \citep{Gruppionis13,Magnelli2013,Marchetti2016,Kilerci2018} in order to have a larger redshift and SFR coverage. We convert the TIR luminosities into SFRs following \citet{Kennicutt2012}:

\begin{eqnarray}
\label{eq_A20}
{\rm Log_{10} (SFR_{IR}) = Log_{10} (L_{TIR}) - 43.41}.
\end{eqnarray}

We note that \citet{Katsianis2020} using radiative transfer/cosmological simulations demonstrated that calibrations that invlove IR wavelengths can overestimate the derived SFRs by 0.5 dex, especially at ${\rm z > 1}$\footnote{The analysis involved galaxies with ${\rm SFRs}$ of ${\rm 0.3 - 100}$ ${\rm M_{\odot}/yr}$  and ${\rm M_{\star}}$ of ${\rm 10^{8.5}- 10^{11}}$ ${\rm M_{\odot}}$ at $z \sim 1-4$. Objects with higher SFRs (and conclusively IR bright objects) were not included since we relied on the EAGLE cosmological simulation which has been found to suffer from a scarcity of SFR/IR bright objects \citep{Cowley2019,Wang2019}}. In addition, \citet{Martis2019} using observations of the UltraVISTA  DR3  photometry  and Herschel PACS-SPIRE  data  demonstrated that commonly adopted relations to derive SFRs from the observed 24${\rm \mu m}$ are found to overestimate the SFR by a factor of 3-5.

\begin{figure*}
\gridline{\fig{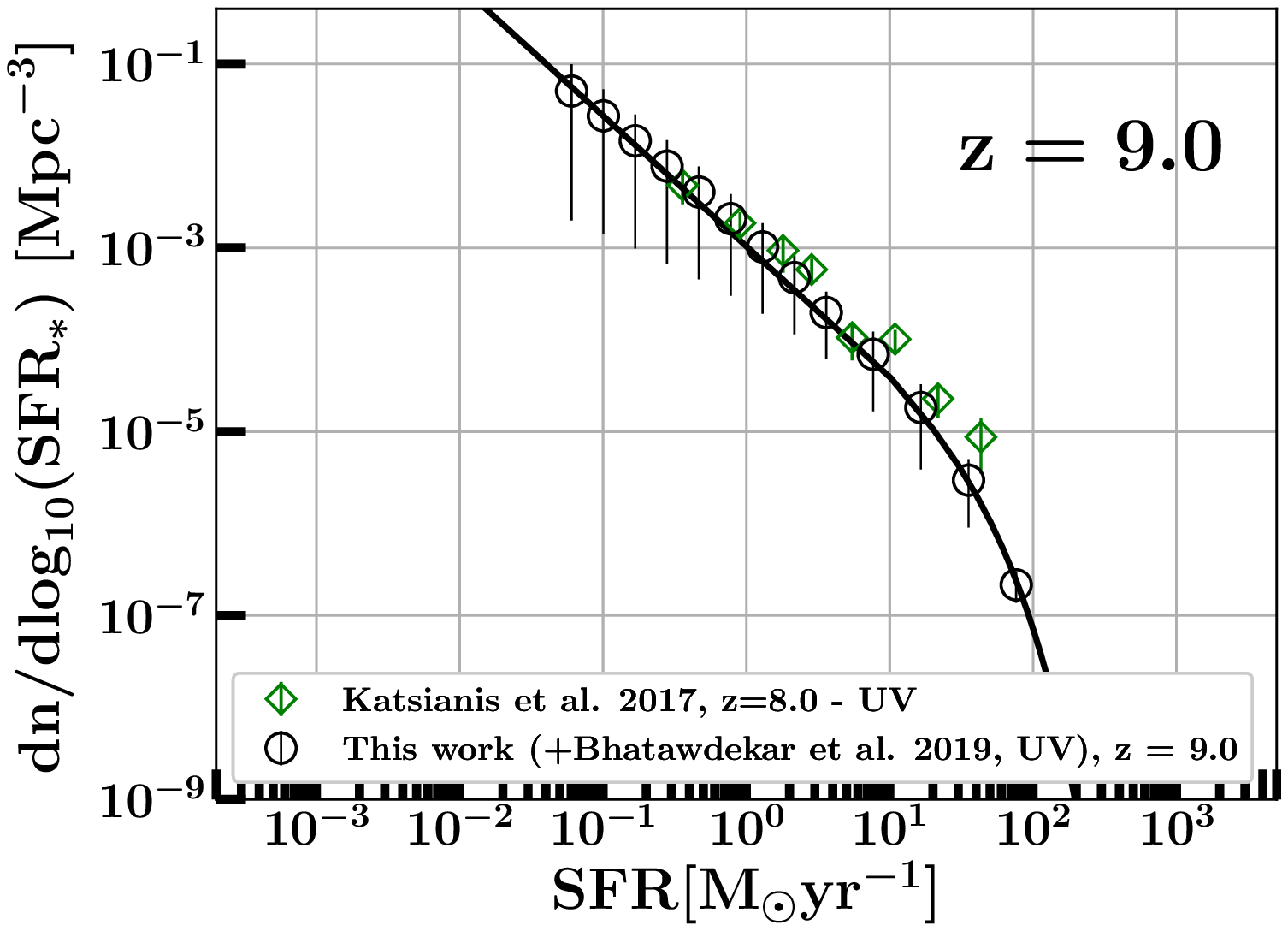}{0.3\textwidth}{}
          \fig{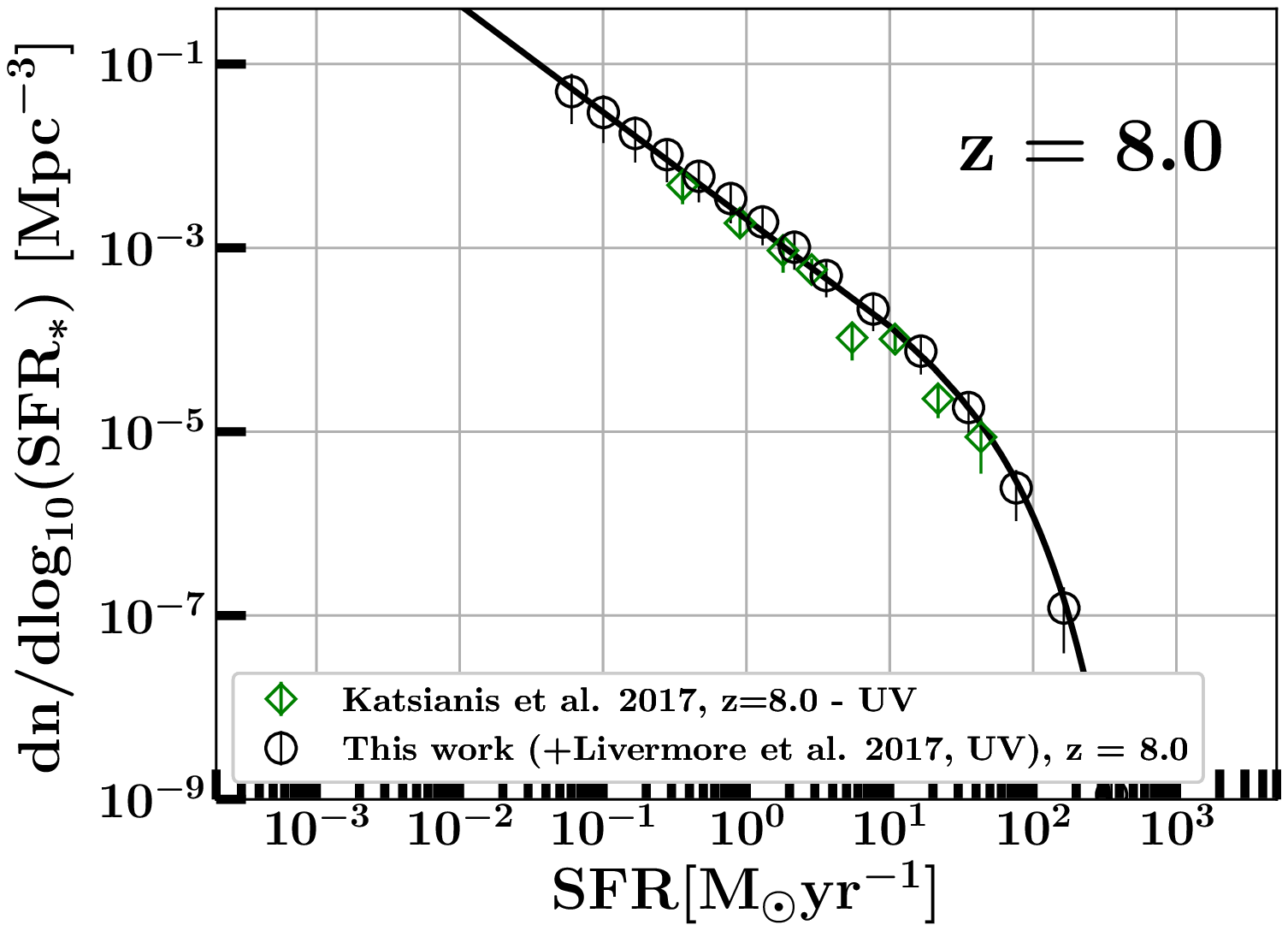}{0.3\textwidth}{}
          \fig{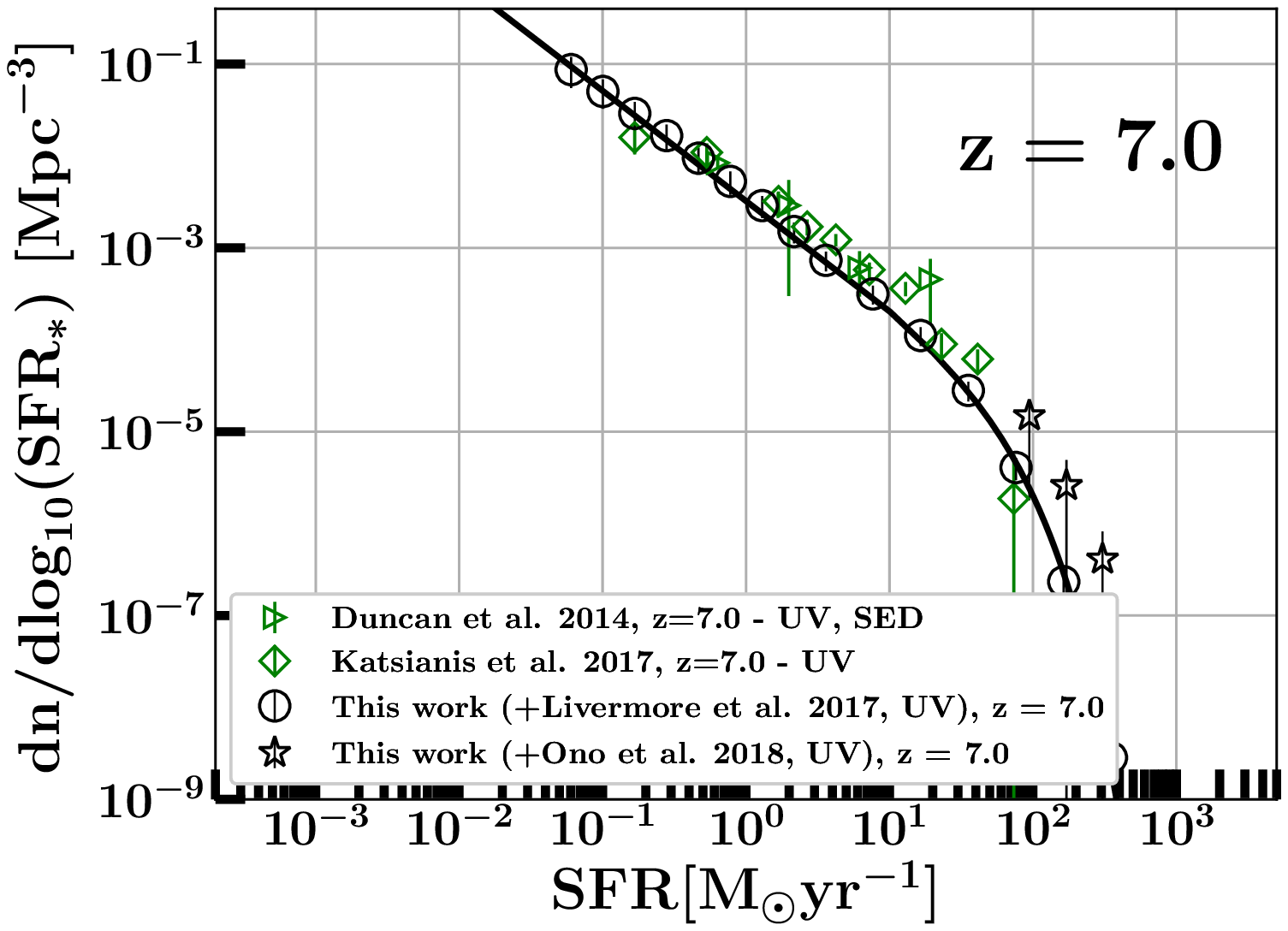}{0.3\textwidth}{}
}
\vspace{-1.54cm}
\gridline{\fig{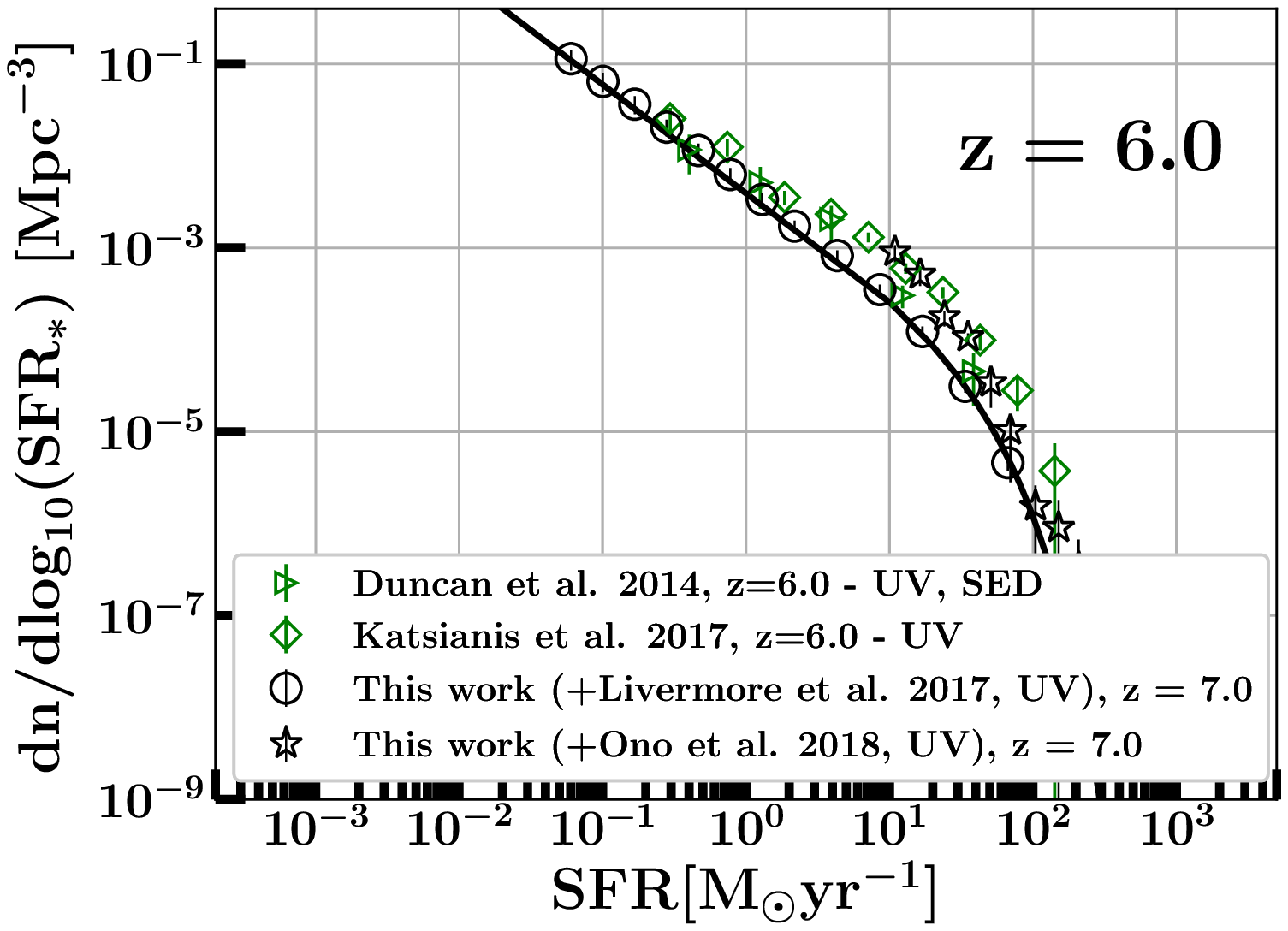}{0.3\textwidth}{}
  \fig{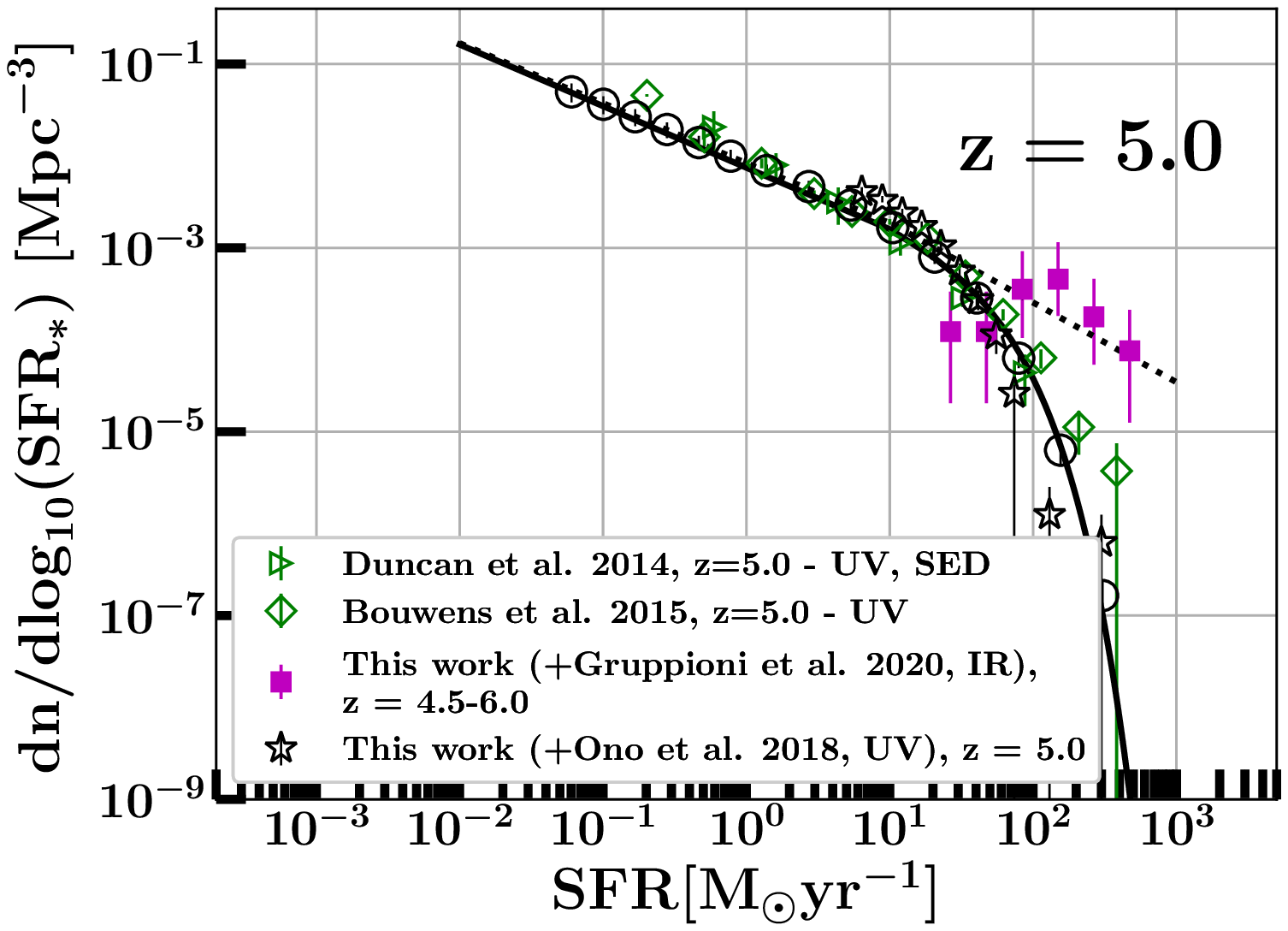}{0.3\textwidth}{}
  \fig{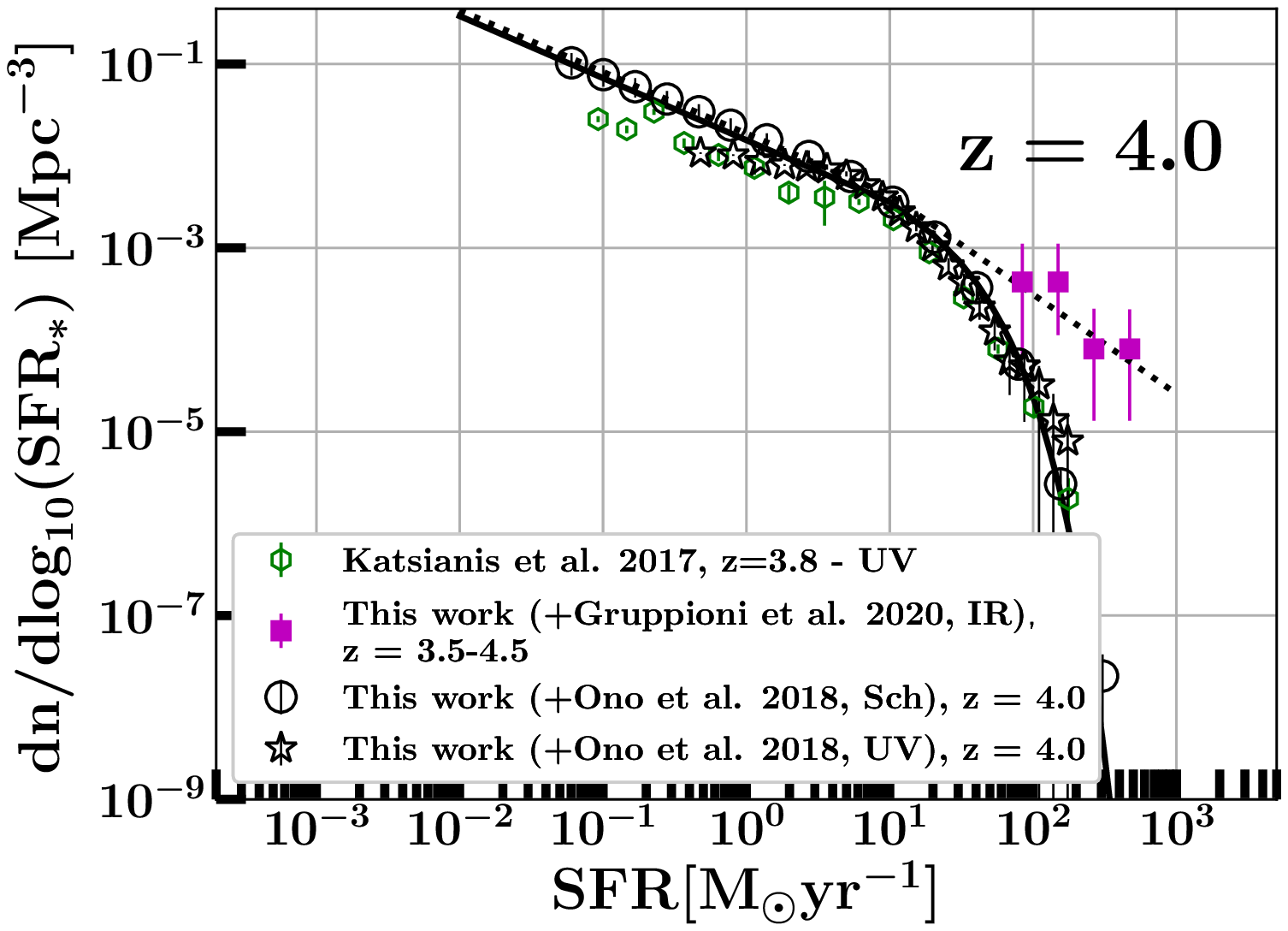}{0.3\textwidth}{}
}
\vspace{-1.54cm}
\gridline{\fig{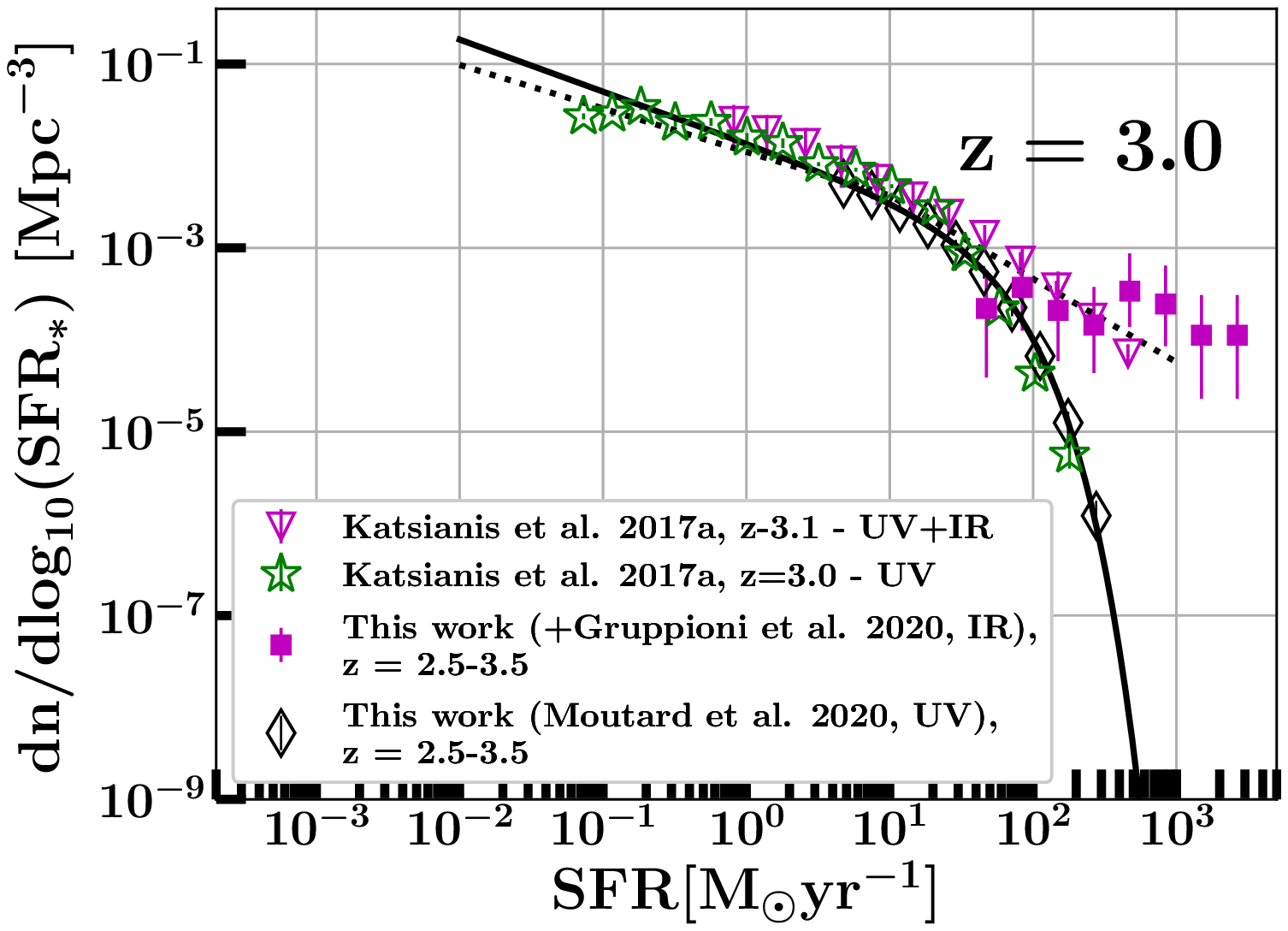}{0.3\textwidth}{}
  \fig{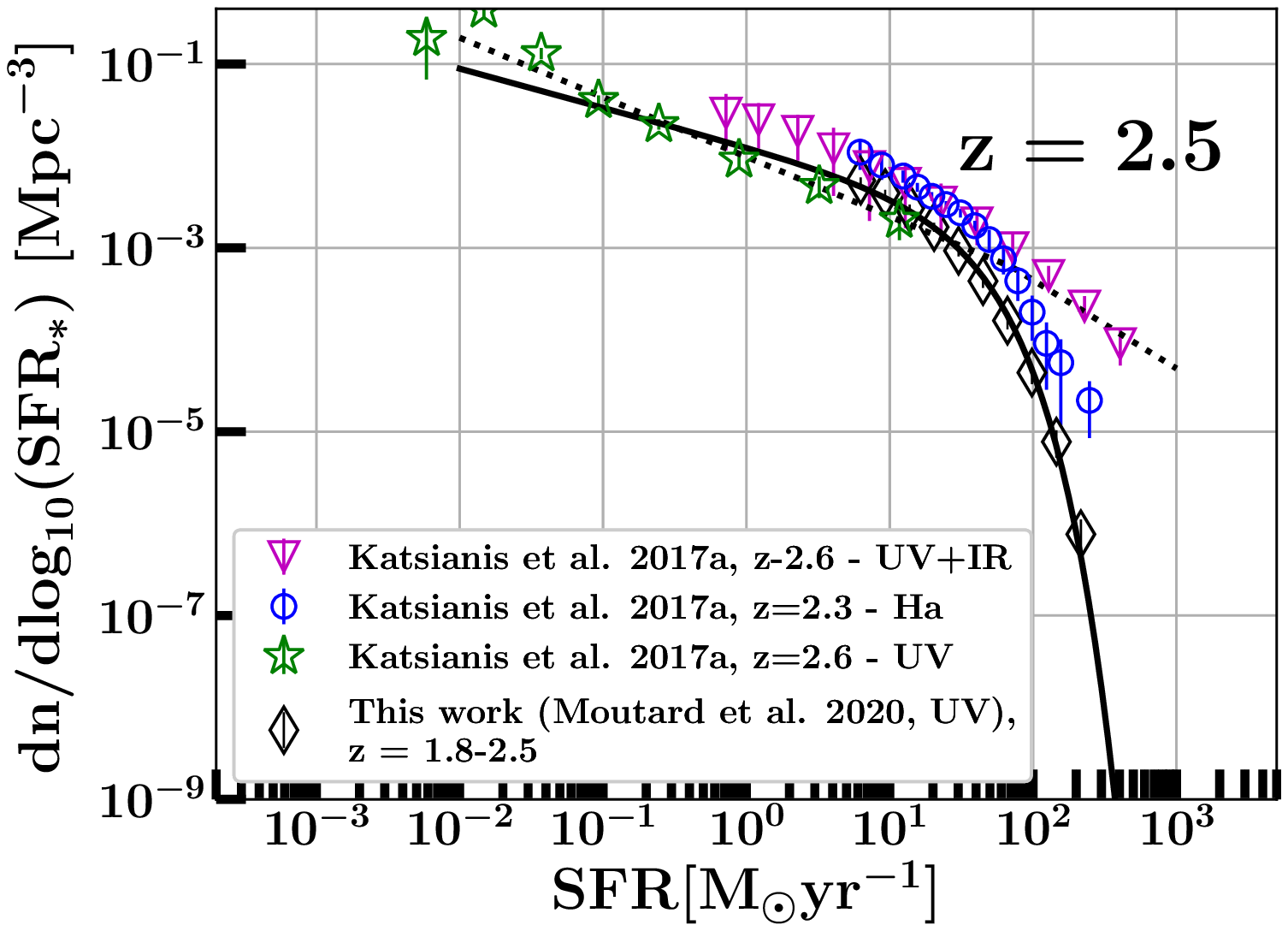}{0.3\textwidth}{}
  \fig{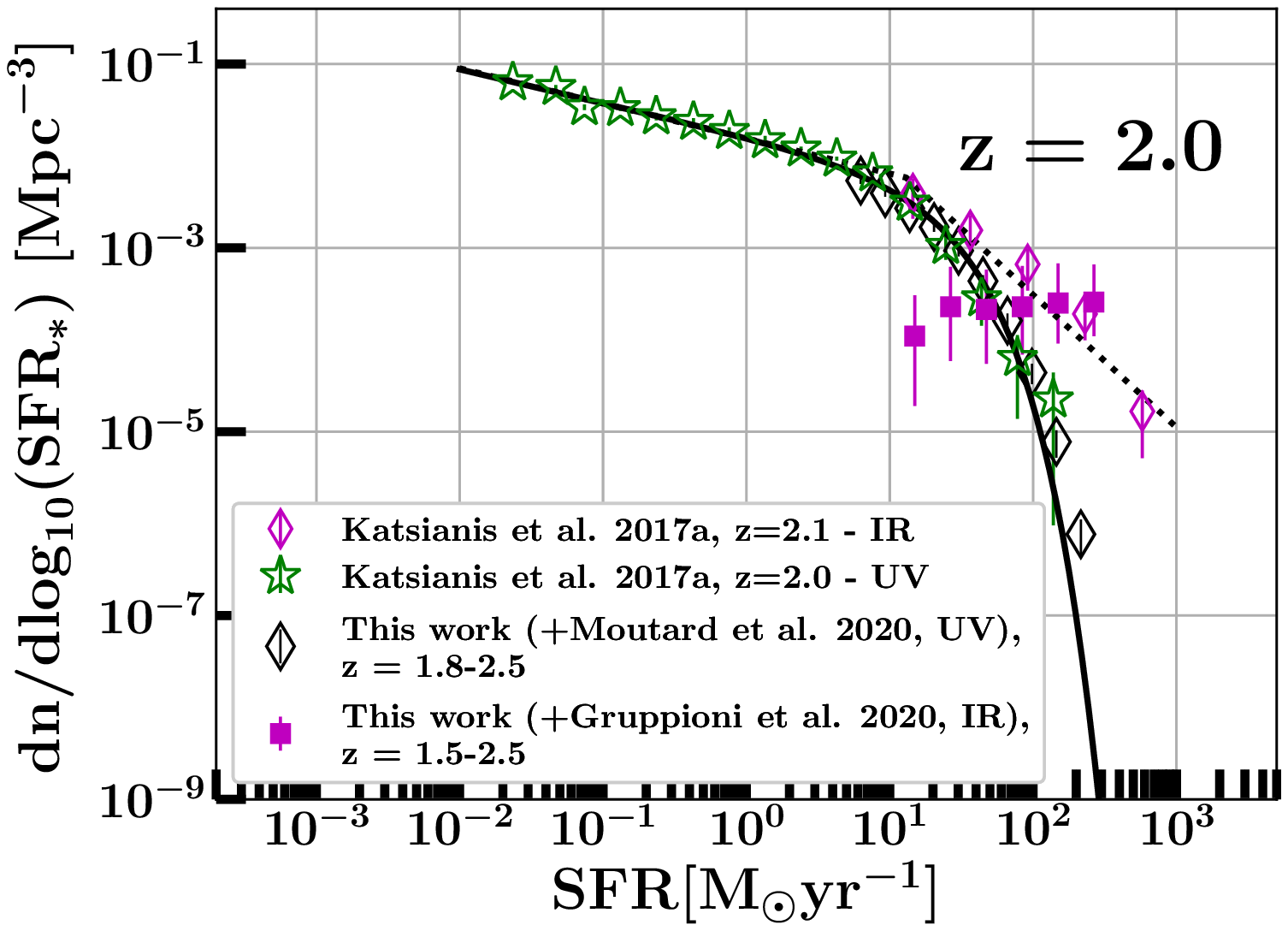}{0.3\textwidth}{}
}
\vspace{-1.54cm}
\gridline{\fig{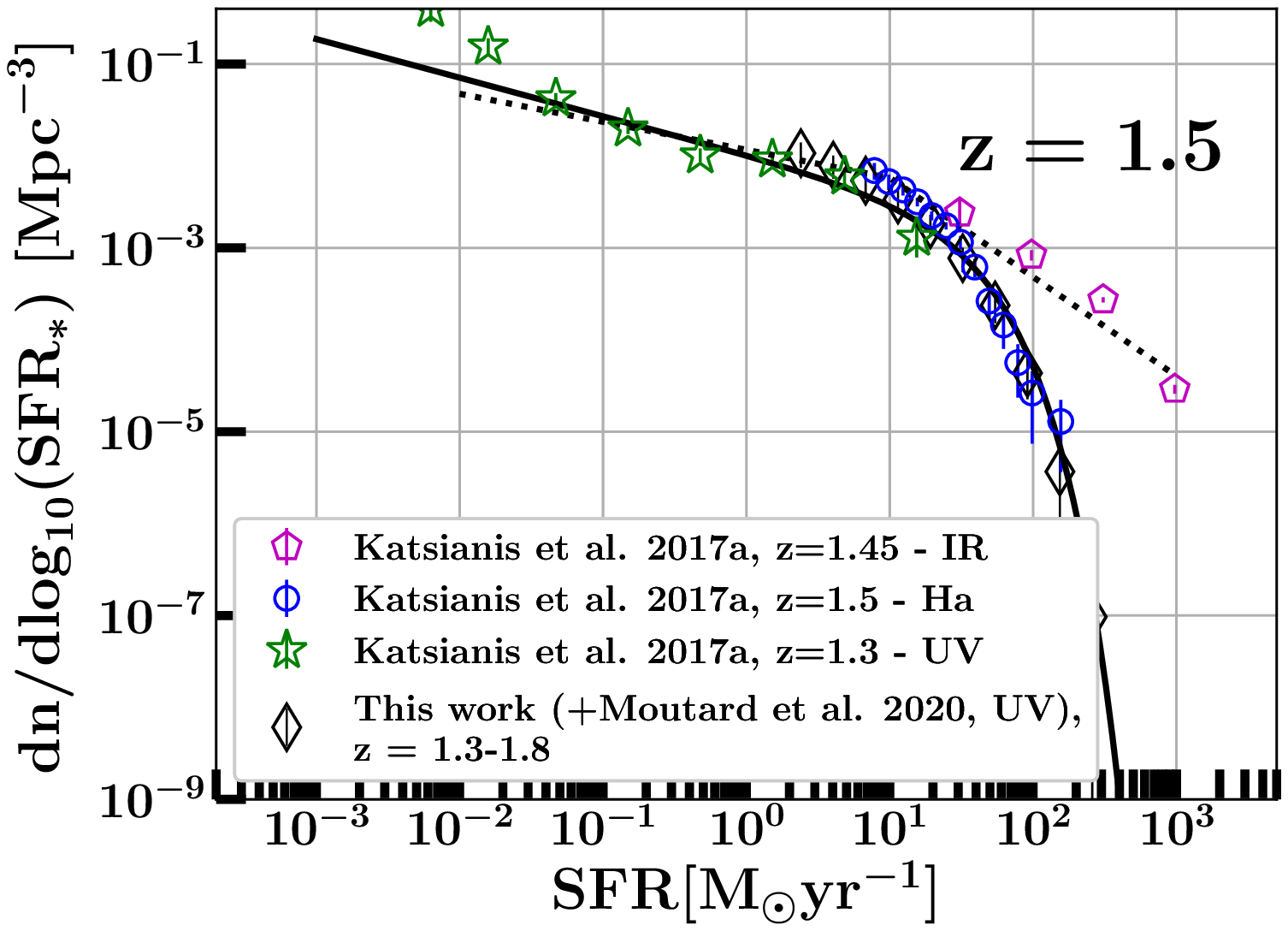}{0.3\textwidth}{}
  \fig{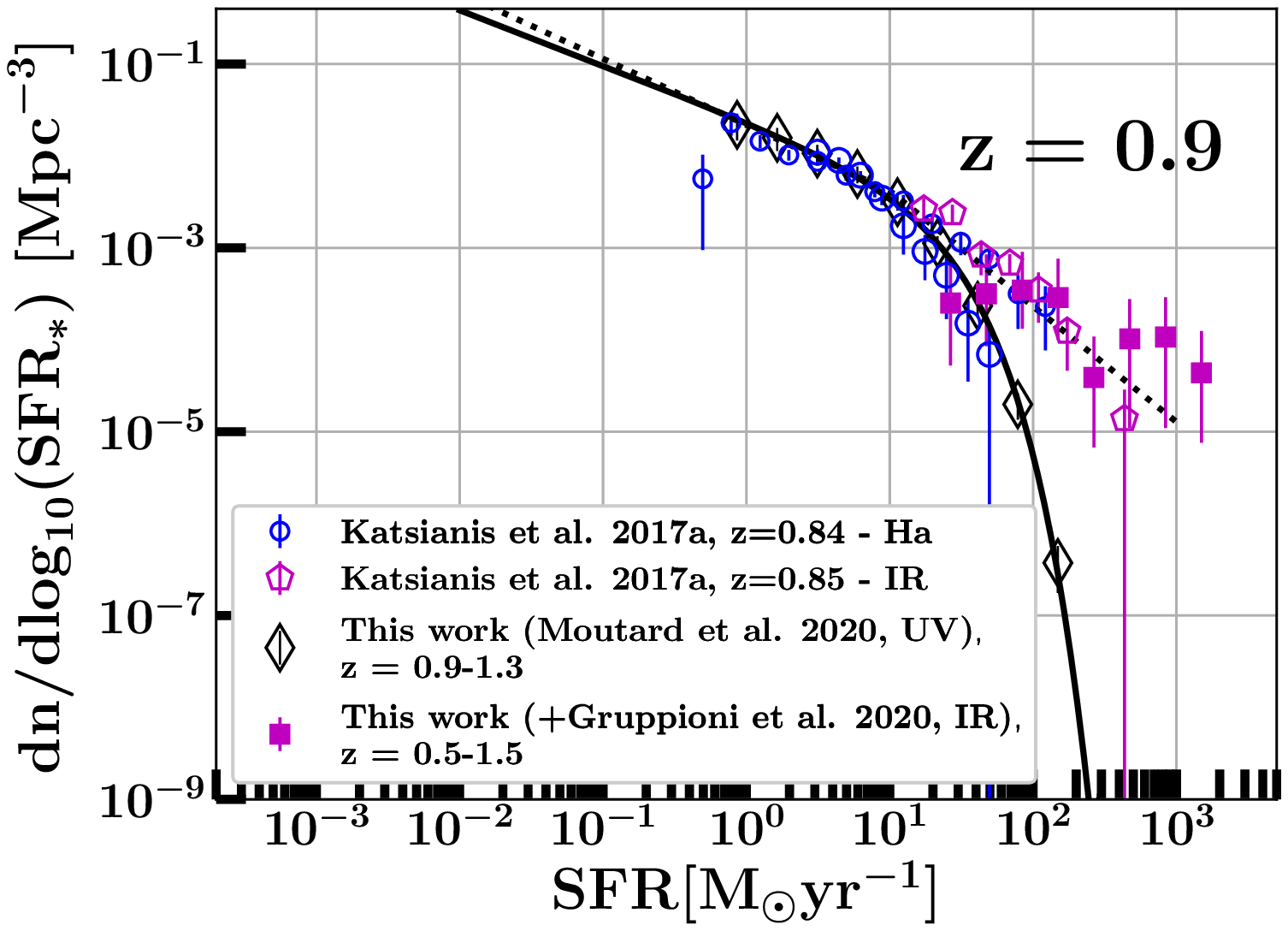}{0.3\textwidth}{}
  \fig{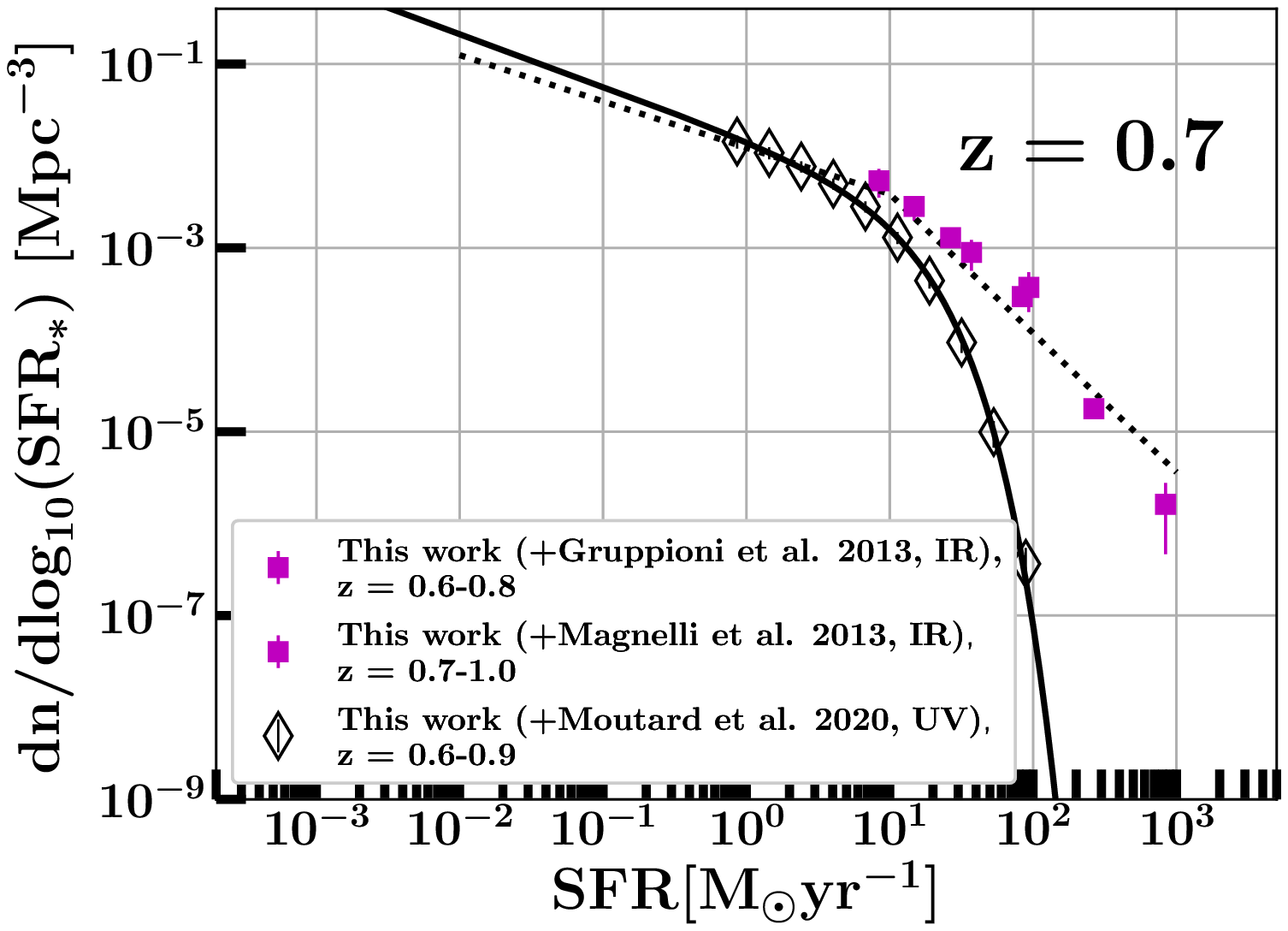}{0.3\textwidth}{}
}
\vspace{-1.54cm}
\gridline{\fig{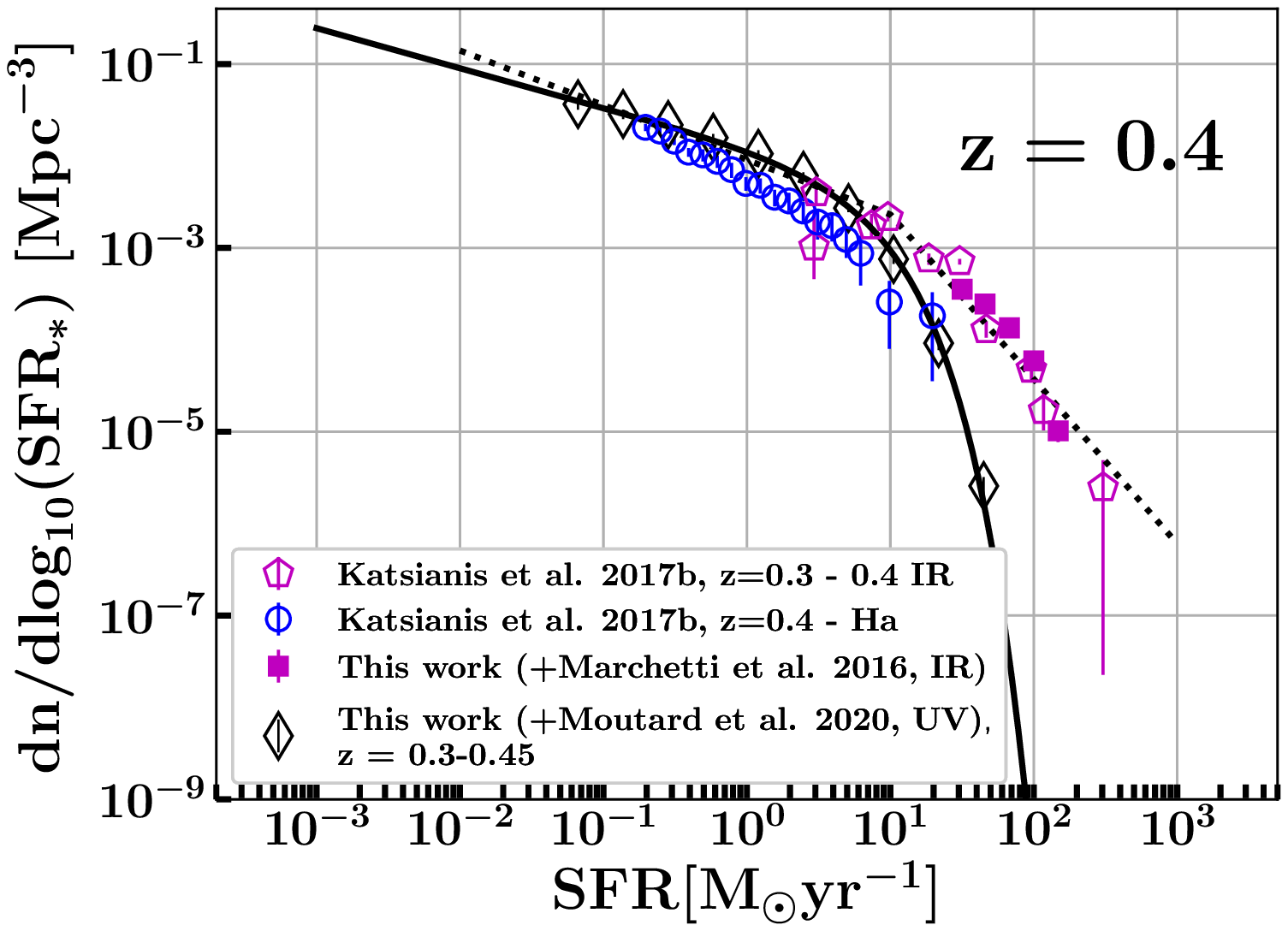}{0.3\textwidth}{}
  \fig{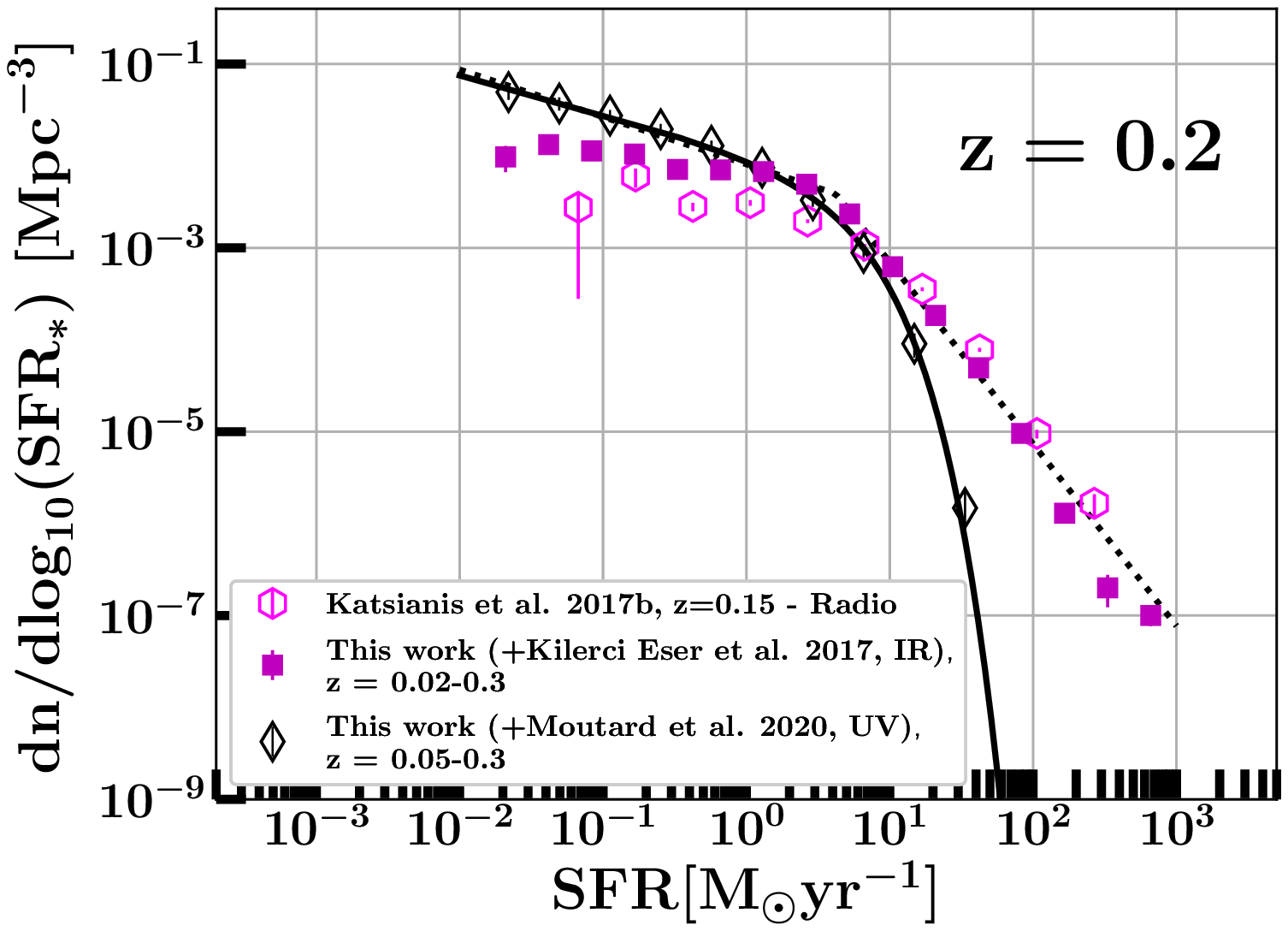}{0.3\textwidth}{}
  \fig{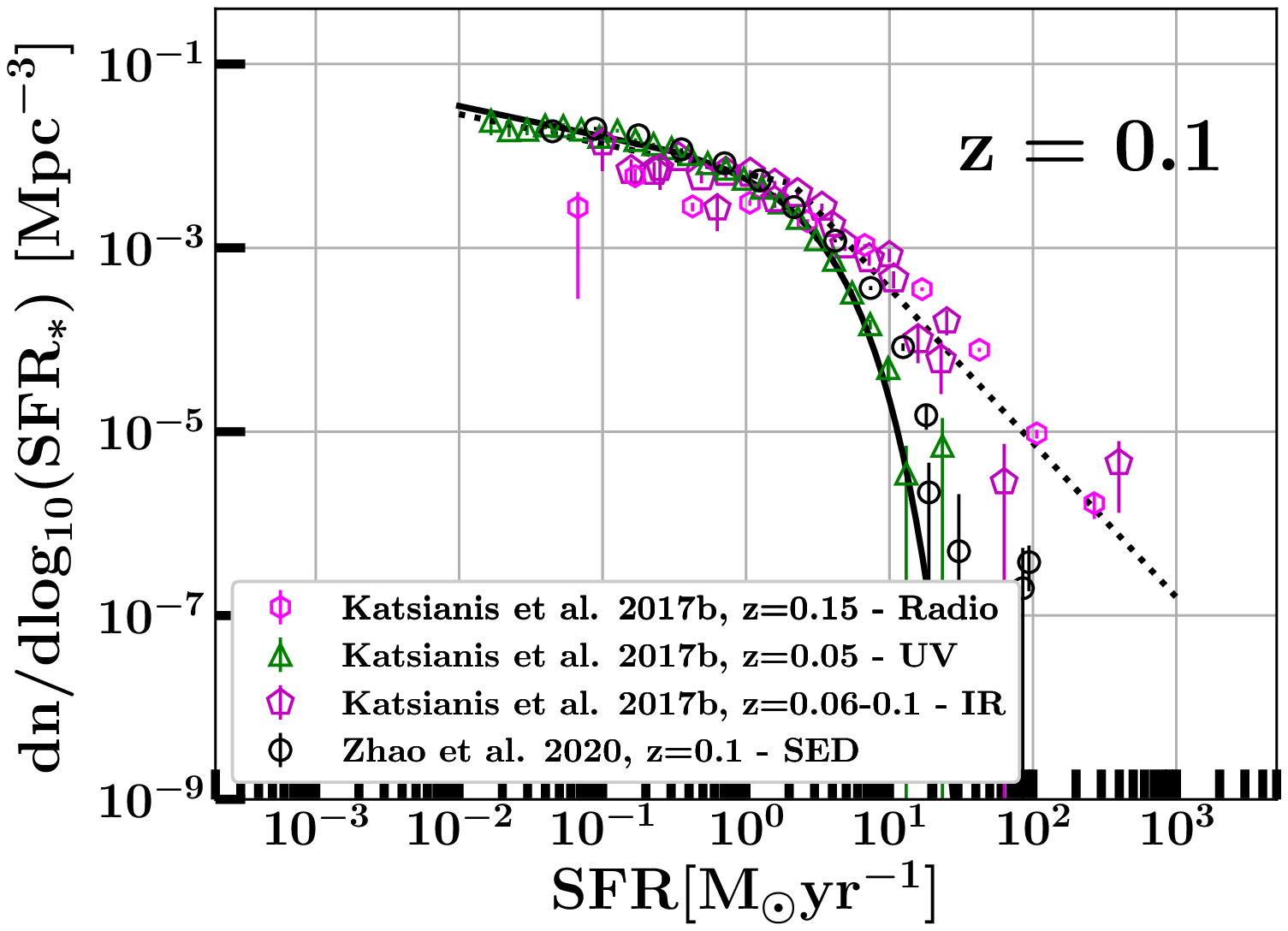}{0.3\textwidth}{}
}
\vspace{-1.24cm}
\caption{The evolution of the star formation rate function obtained from CLAUDS+HCS UV observations \citep[black filled diamonds, ][]{Moutard2020} and a compilation of IR studies \citep[black filled squares, ][]{Gruppionis13,Magnelli2013,Marchetti2016,Kilerci2018,Gruppioni2020}. Alongside we plot the results from \citet{Katsianis2016} and \citet{Katsianis2017}, converted to a Chabrier IMF \citep{chabrier03} and the \citet{Kennicutt2012} calibration when required. The solid black lines represent the Schechter fits of the ${\rm UV_{corr}}$-SFRFs, while the black dotted line is a double-power law fit to the high star forming end obtained from IR studies \citep{Gruppionis13,Magnelli2013,Marchetti2016,Kilerci2018} and the low star forming end probed by the UV studies \citep{Katsianis2016,Katsianis2017,Moutard2020}).
\label{fig:SFRIR1}}
\end{figure*}

\begin{deluxetable}{ccccccccc}
  \tablecaption{The parameters of the ${\rm UV_{corr}}$ SFRFs (black solid lines of Fig. \ref{fig:SFRIR1}), UV+IR SFRFs (black dotted lines of Fig. \ref{fig:SFRIR1}),  ${\rm CSFRD_{UV, corr}}$ (black dashed line of top right panel of Fig. \ref{fig:SFRIR1EEE}) and ${\rm CSFRD_{UV+IR}}$ (purple dashed line of top right panel of Fig. \ref{fig:SFRIR1EEE}) \label{table22}}
  \tablehead{
\colhead{UV-SFRF, redshift} & \colhead{$\Phi_{\star, Sch}$} & \colhead{$\Phi_{\star, unc, Sch}$} & \colhead{$\alpha$} & \colhead{$\alpha_{unc}$} & \colhead{$SFR_{\star}$}  & \colhead{$SFR_{\star, unc}$} & \colhead{$CSFRD$} & \colhead{$CSFRD_{unc}$} }
  \startdata
 8.0  & 0.18 & 0.07 & -2.23 & 0.08 & 23.01 & 3.23  & 0.009 & 0.002 \\
 7.0  & 0.39 & 0.01 & -2.10 & 0.01 & 33.17 & 3.63  & 0.015 & 0.003 \\
 6.0  & 0.26 & 0.02 & -2.16 & 0.08 & 40.05 & 1.94  & 0.018 & 0.002 \\
 5.0  & 3.30 & 0.08 & -1.68 & 0.03 & 41.04 & 5.68  & 0.034 & 0.004 \\
 4.0  & 11.82 & 0.08 & -1.60 & 0.02 & 22.12 & 0.85  & 0.055 & 0.008 \\
 3.0  & 7.24  & 1.41 & -1.56 & 0.08 & 42.89 & 4.00 & 0.061 & 0.007 \\
 2.6  & 13.68 &  1.25 & -1.42 & 0.02 & 26.91 & 1.35 & 0.056 & 0.006 \\
 2.15 & 23.74 &  1.04 & -1.37 & 0.05 & 19.78 & 0.82 & 0.066 & 0.010  \\
 1.55 & 12.31 &  4.5 & -1.61 & 0.16 & 22.72 & 3.65 & 0.061 & 0.011 \\
 1.1  & 18.23 &  4.22 & -1.60 & 0.2 & 17.88 & 1.66 & 0.069 & 0.008 \\
 0.7  & 17.4 &  3.0 & -1.57 & 0.08 & 10.4  & 0.8 & 0.037 & 0.004 \\
 0.37 & 25.9 &  2.63 & -1.43 & 0.03 & 6.10 & 0.3 & 0.024 & 0.002 \\
 0.18 & 26.5 &  4.0 & -1.42 & 0.04 & 4.068 & 0.36 & 0.016 & 0.003 \\
&  & & & & & & & \\ \hline
\hline
UV+IR, redshift  & $\Phi_{\star, double}$ & $\Phi_{\star, unc, double}$ & $\alpha_{1}$ & $\alpha_{1, unc}$ & $\alpha_{2}$ & $\alpha_{2, unc}$ & $SFR_{break}$ &  $ CSFRD$ \\
\hline
 5.25 & 0.019 & 0.002 & 1.65 & 0.03 & 1.86 & 0.14  & 10 & 0.073  $\pm$ 0.014 \\
 4.0  & 0.037 & 0.004 & 1.67 & 0.04 & 2.08 & 0.15  & 10 & 0.106  $\pm$ 0.022 \\
 3.0  & 0.038  & 0.025 & 1.47 & 0.19 & 1.91 & 0.23 & 10 & 0.123  $\pm$ 0.015 \\
 2.15 & 0.064 &  0.002 & 1.38 & 0.01 & 2.36 & 0.11 & 13 & 0.107  $\pm$ 0.008  \\
 1.4  & 0.058 &  0.007 & 1.30 & 0.04 & 2.07 & 0.04 & 10 & 0.133  $\pm$ 0.011 \\
 1.0  & 0.041 &  0.006 & 1.74 & 0.11 & 2.26 & 0.11 & 10 & 0.104  $\pm$ 0.027  \\
 0.7  & 0.038 &  0.004 & 1.54 & 0.07 & 2.53 & 0.05 & 10 & 0.062  $\pm$ 0.010 \\
 0.35 & 0.023 &  0.002 & 1.59 & 0.02 & 2.79 & 0.05 & 10  & 0.036 $\pm$ 0.003 \\
 0.16 & 0.016 &  0.001 & 1.51 & 0.02 & 2.98 & 0.02 &  4.2 & 0.021 $\pm$ 0.004 \\
 0.08 &  0.0105 &  0.0003 & 1.33 & 0.01 & 2.68 & 0.07 & 2.1 & 0.013 $\pm$ 0.001 \\
&  & & & & & & & \\
\enddata
\tablecomments{The normalization $\Phi_{\star, Sch}$ and its uncertainty $\Phi_{\star, Sch}$ are given in units of ${\rm 10^{-4}}$ ${\rm Mpc^{-3}}$, while the $\Phi_{\star, double}$ and its uncertainty $\Phi_{\star, double}$ are given in units of ${\rm 10^{-4}}$ ${\rm Mpc^{-3} \, dex^{-1}}$. The characteristic $SFR_{\star}$ and power-law break $SFR_{break}$ are given in ${\rm M_{\odot} \, yr^{-1} }$, while the CSFRD is given in units of ${\rm M_{\odot} \, yr^{-1} \, Mpc^{-3}}$. The integrations of the CSFRDs occurs between 0.01 to 1000  ${\rm M_{\odot} \, yr^{-1} }$ consistently for all redshifts since these limits do not require extending our SFRFs to regimes where there are no data.}
\end{deluxetable}

\begin{figure*}
  \centering
\includegraphics[scale=0.47]{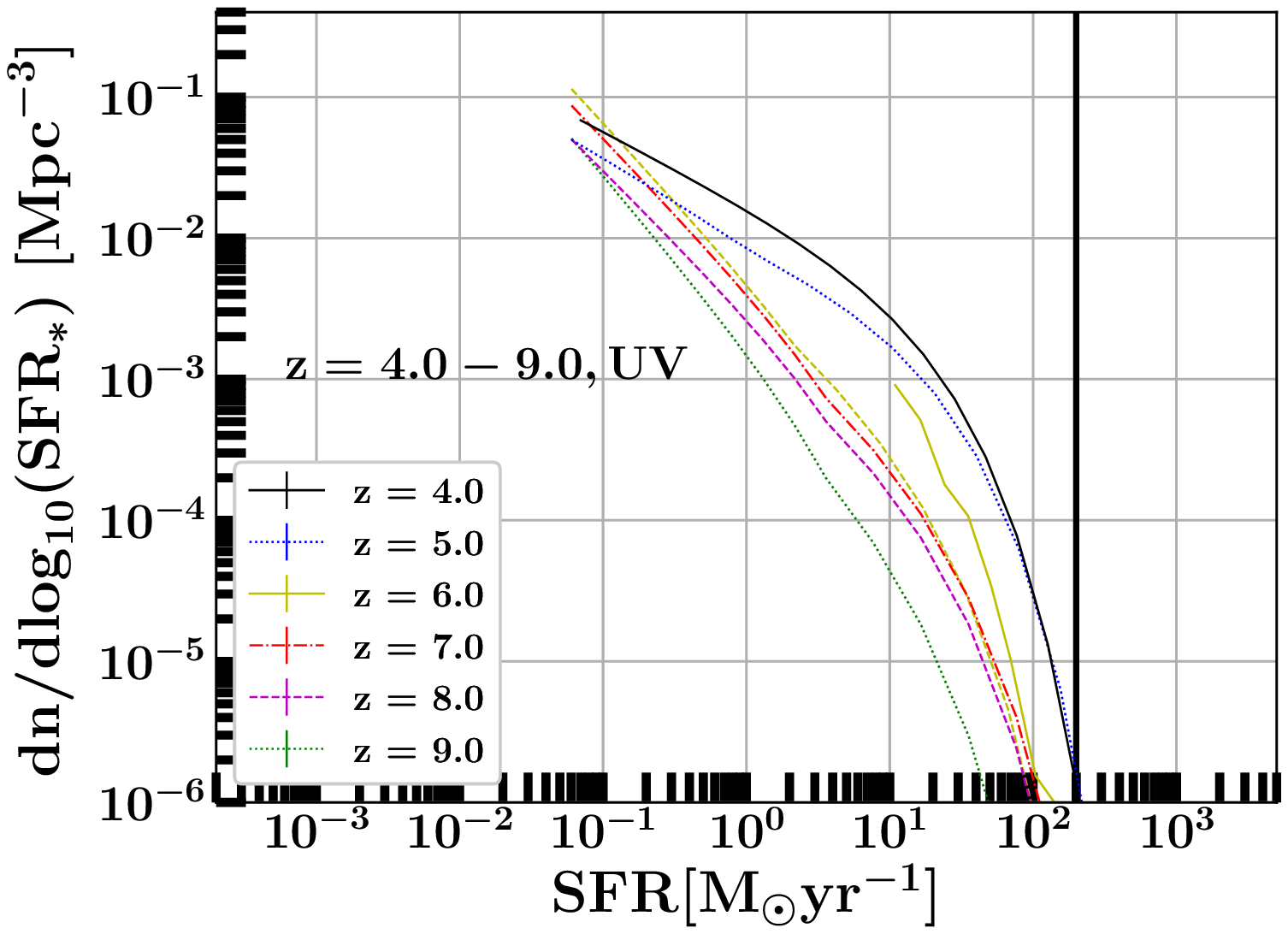}
\vspace{-0.50cm}
\includegraphics[scale=0.47]{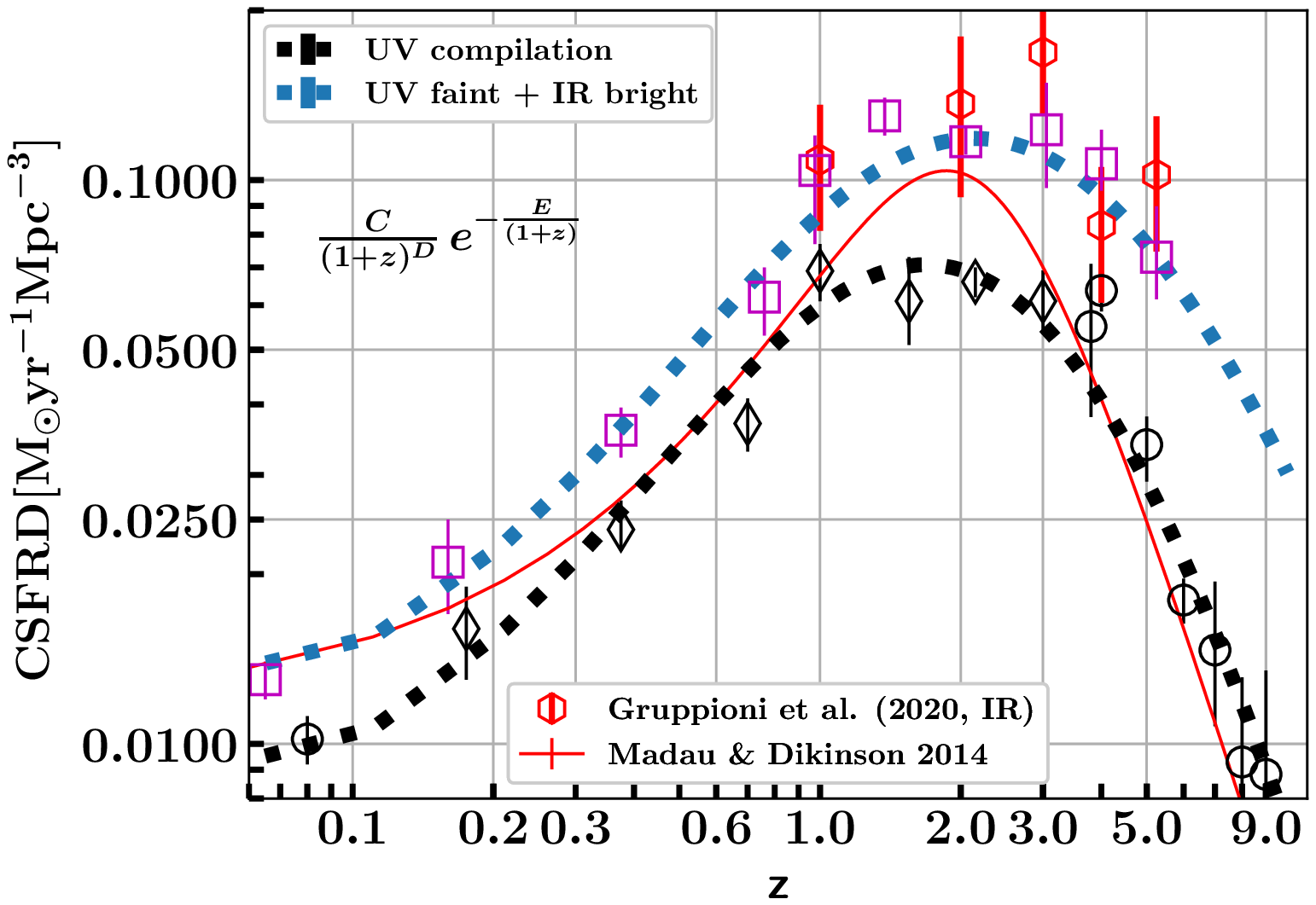}
\includegraphics[scale=0.47]{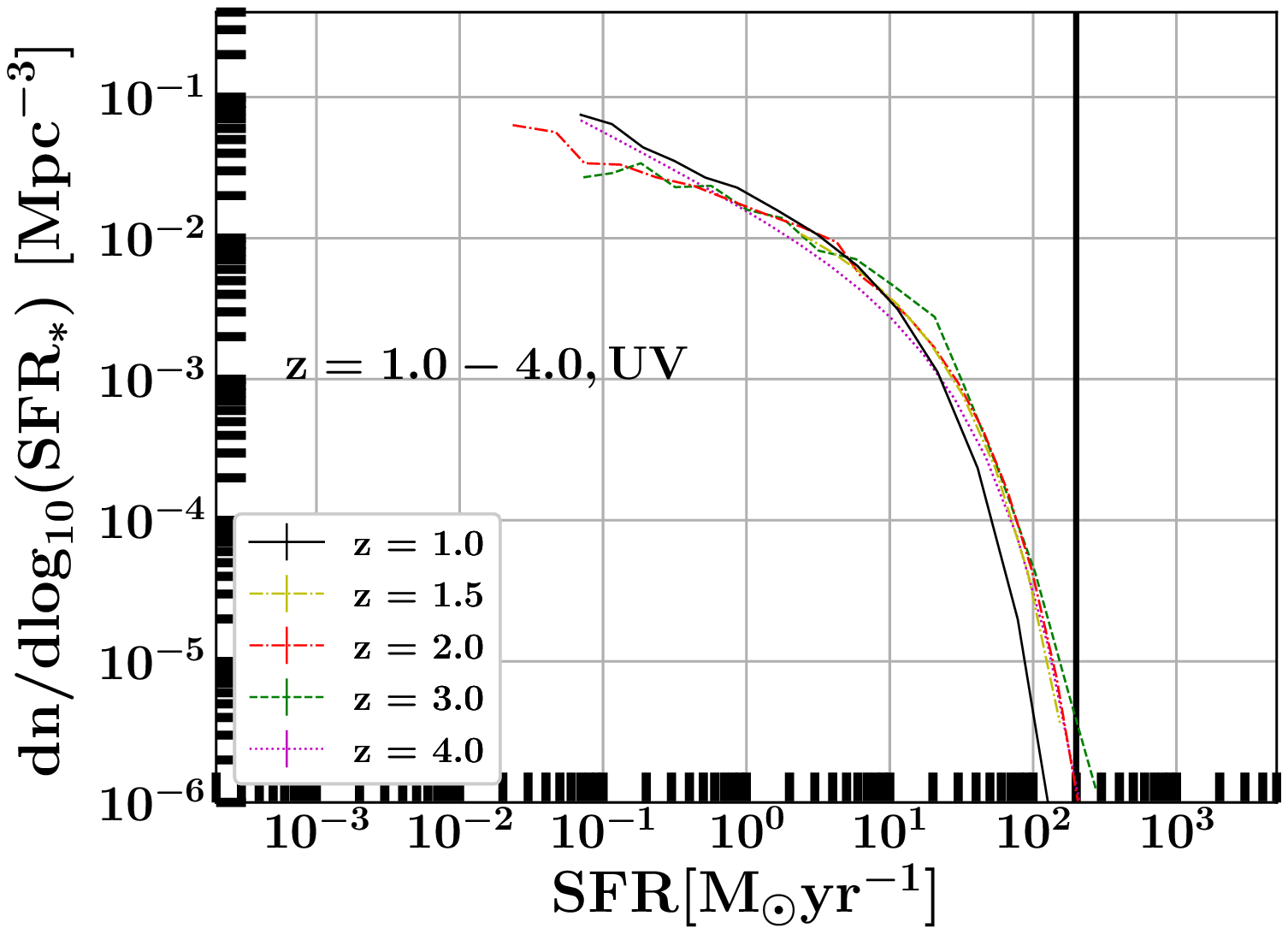}
\vspace{-0.50cm}
\includegraphics[scale=0.47]{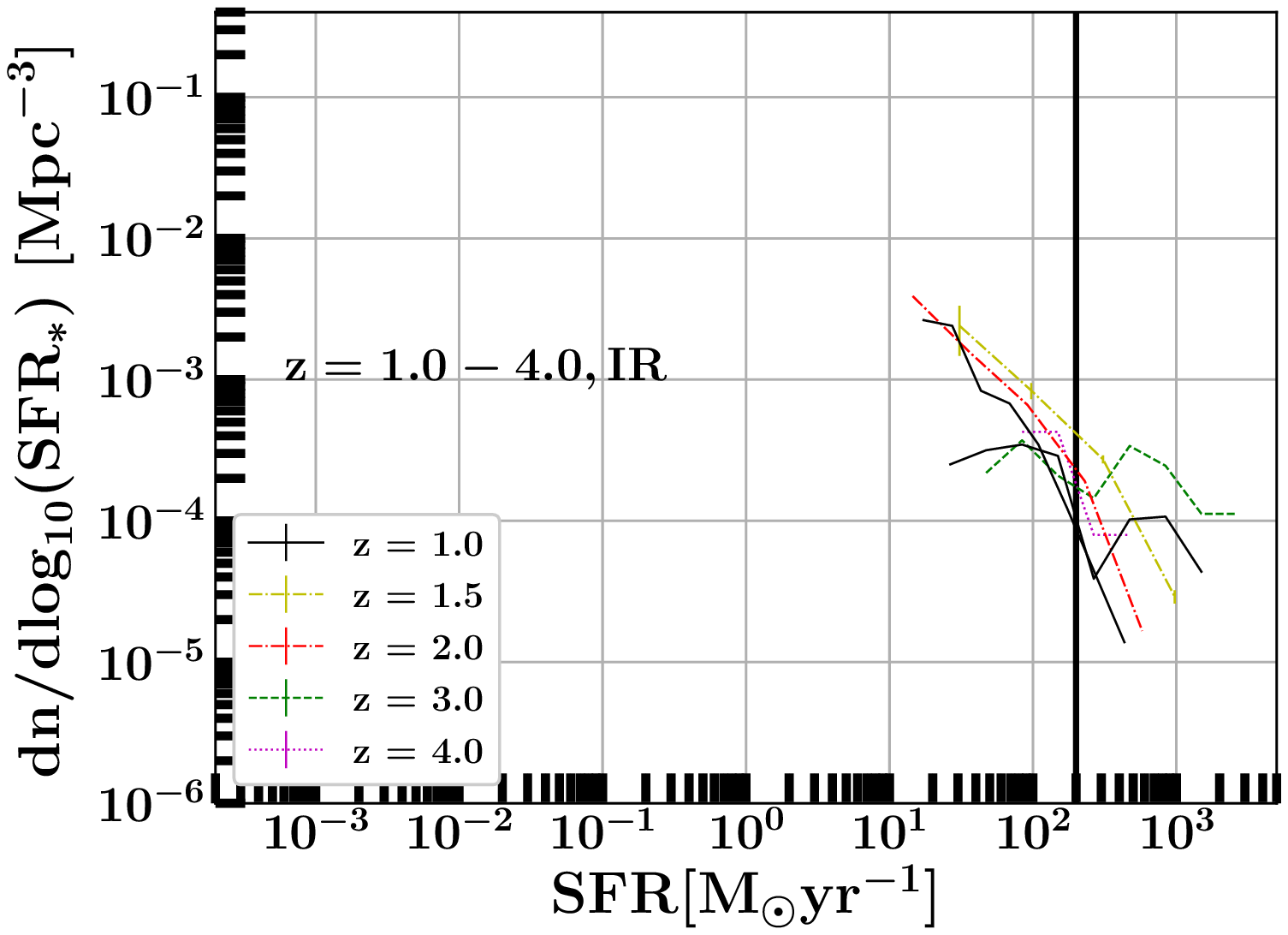}
\includegraphics[scale=0.47]{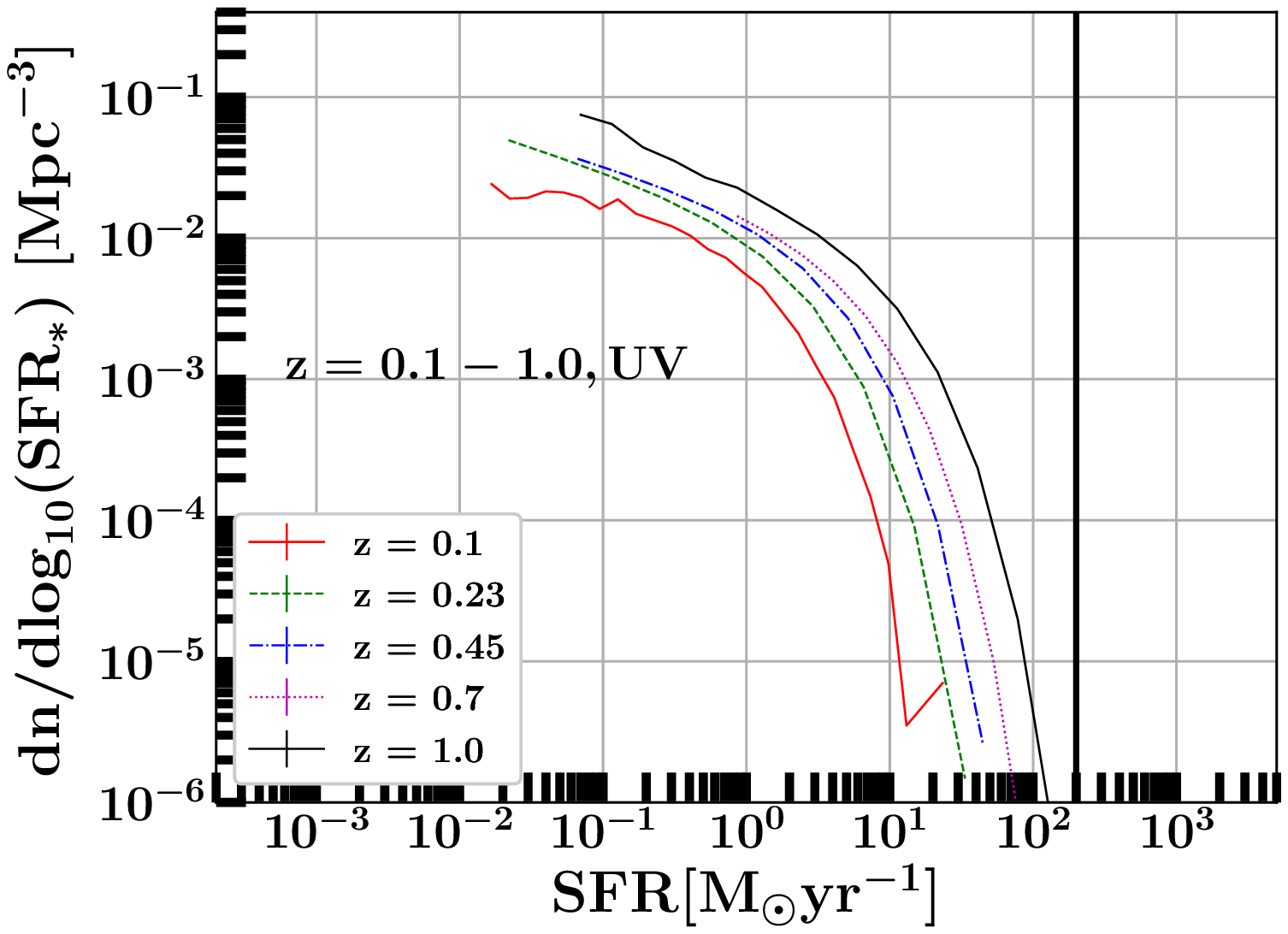}
\includegraphics[scale=0.47]{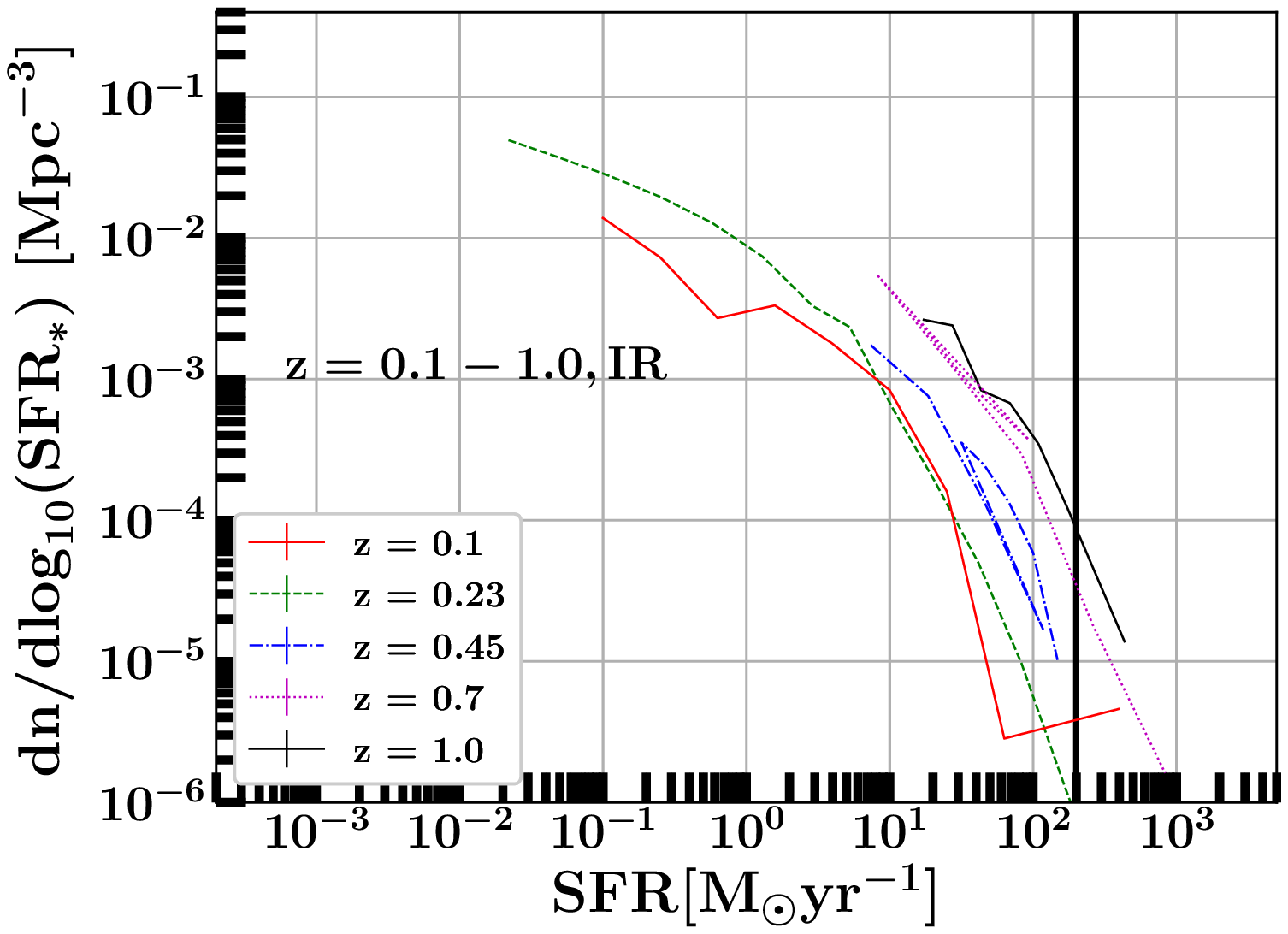}
\vspace{-0.50cm}
\caption{Top left panel, middle left panel, bottom left panel: The evolution of the UV SFRF at ${\rm z \sim 4-9}$, ${\rm z \sim 1-4}$ and ${\rm z \sim 0-1}$, respectively. Top right panel: The evolution of the CSFRD from the integration of the SFRFs present in Fig. \ref{fig:SFRIR1}. Middle right panel, bottom right panel: evolution of the IR SFRF at ${\rm z \sim 1-4}$ and ${\rm z \sim 0-1}$, respectively. We remind that we adopt a Chabrier IMF \citep{chabrier03} and \citet{Kennicutt2012} relations.}
\label{fig:SFRIR1EEE}
\end{figure*}

\subsection{The ``observed'' star formation rate function}
\label{SFRF}

In this subsection we present the evolution of the star formation rate function at ${\rm z \sim 0-9}$. Alongside we plot the results from \citet{Katsianis2016} and \citet{Katsianis2017} represented by the open blue circles (H$\alpha$), open magenta pentagons/triangles (IR), open green stars/diamonds (UV+dust corrections). In Fig. \ref{fig:SFRIR1} and table \ref{table22} we summarize the results of this work. The evolution of the star formation rate function obtained from the CLAUDS+HCS UV observations \citep{Moutard2020} is represented by the open black diamonds. The open black circles and the open black stars describe the SFRFs obtained from \citet{Livermore2017} and \citet{Ono2018}, respectively. The solid black lines represent a fit for the above UV-SFRFs that is described by a Schechter form. In Fig. \ref{fig:SFRIR1EEE} we present the evolution of the SFRF at 3 different redshift ranges (top left panel - ${\rm z \sim 4-9}$, middle left panel - ${\rm z \sim 1-4}$ and bottom left panel - ${\rm z \sim 0-1}$ describe the UV SFRF evolution). We note the following:

\begin{itemize}
  
\item The UV SFRFs of this work are broadly described by a standard Schechter form (Fig. \ref{fig:SFRIR1}) as follows:

\begin{eqnarray}
\label{eq:schechter}
{\rm \Phi_{SFR} \, dlogSFR = ln(10) \times \Phi_{\star, Sch} \times \left( \frac{SFR}{SFR_{\star}}  \right)^{\alpha+1} \times e^{-\frac{SFR}{SFR_{\star}}}}.
\end{eqnarray}  

The parameters at different redshifts can be found in table \ref{table22}. 

\item there are almost no objects with ${\rm SFR_{UV, Corr}}$ more than ${ \rm \sim \, 100  \, M_{\odot} \, yr^{-1}}$ (Fig. \ref{fig:SFRIR1EEE}). A clear maximum SFR limit is implied. We note that the ${\rm UV_{corr}}$ SFRFs at the high star forming end have been found to be in good agreement with the predictions from  the EAGLE \citep{Katsianis2017} and IllustrisTNG simulations \citep{Zhaoka2020} but lower with respect other models like Simba \citep{dave2019,Lovell2021b} which are able to produce more massive and high star forming galaxies,

\item the SFRF increase in normalization and becomes less steep with time (top left panel of Fig. \ref{fig:SFRIR1EEE}) from redshift 9 to 4. The evolution of the SFRF is mostly driven by the emergence of high star forming systems. This results to an increasing CSFRD\footnote{We note that the integration limits for the calculation of the CSFRD from the SFRF are SFR = 0.01 - 1000 ${\rm M_{\odot} \, yr^{-1}}$, consistently for both indicators and all redshifts considered.} (shown in Fig. \ref{fig:SFRIR1EEE} as the black dotted line). However, the SFRF is mostly unchanged from redshift 4 to 1 (shown at the middle left panel of Fig. \ref{fig:SFRIR1EEE}) and this results to a CSFRD that is constant at this era. There is no peak at ${\rm z \sim 2}$ in contrast with \citet{Madau2014}. The indicator suggests that there is instead a plateau that remains constant at ${\rm z \sim 1-4}$ in agreement with \citet{Moutard2020} and \citet{Gruppioni2020}. At ${\rm z \sim 1}$ to 0 there is a uniform decrement of the SFRF that occurs similarly at all star formation rate bins that causes a decrement to the CSFRD  (shown at the top left panel of Fig. \ref{fig:SFRIR1EEE}). We note that according to the above ${\rm UV_{corr}}$ SFRFs there is not a more aggressive quenching for the high or low star forming objects and this can be interesting for studies that focus on galaxy feedback.
  
\end{itemize}

The SFRFs obtained from the compilation of IR studies \citep{Gruppionis13,Magnelli2013,Marchetti2016,Kilerci2018,Gruppioni2020} are represented by the magenta filled squares. We see that the above IR SFRFs are able to probe only the high star forming end (${\rm SFR > 10 M_{\odot} \, yr^{-1}}$) at high redshifts so we combine them with the UV SFRF for the low star forming end\footnote{UV SFRs, besides being the only constraint at low star formation rates are found to be ``reliable'' for ${\rm SFR < 10 \, \, M_{\odot} \, yr^{-1}}$ objects at all redshifts, see subsection introduction} in order to construct an analytical form described by the black dotted line of Fig. \ref{fig:SFRIR1}. We suggest that ambiguously extending the IR SFRFs to lower SFRs is problematic as discussed in the introduction since there are not data from the indicator at the low star forming end. We note that by doing so we essentially compare the high star forming ends of the distributions and this describes a lower limit for the tension between our ${\rm UV_{corr}}$ and UV+IR SFRFs when the analytic forms are compared. In Fig. \ref{fig:SFRIR1EEE} we present the evolution of the SFRF at 2 different redshift eras ( middle right panel - ${\rm z \sim 1-4}$ and bottom right panel - ${\rm z \sim 0-1}$ describe the IR SFRF evolution). We note the following:
\begin{itemize}
\item The UV+IR SFRFs of this work are described by a double-power law (Fig. \ref{fig:SFRIR1})
  \footnote{We note that the IR LF is typically described by double-power laws \citep{WangRobinso2010,Goto2011,Patel13,Kilerci2018} but other authors \citep{Cirasuolo2007,Bell2007,Koprowski2017} favor a Schechter form.} instead of a Schechter form which is described from the following form:,
\begin{eqnarray}
\label{eq:doublepower}
      {\rm \Phi_{SFR} \, dlogSFR = \frac{\Phi_{\star, double}}{SFR_{break}} \, (\frac{SFR}{SFR_{break}})^{1-\alpha_{1}}}  \, {\rm [SFR < SFR_{break}]},  \nonumber \\ 
      {\rm \Phi_{SFR} \, dlogSFR = \frac{\Phi_{\star, double}}{SFR_{break}} \, (\frac{SFR}{SFR_{break}})^{1-\alpha_{2}}}  \, {\rm [SFR > SFR_{break}]}. 
\end{eqnarray}
The parameters at different redshifts can be found in table \ref{table22}. 
\item in contrast to the ${ \rm UV_{corr}}$ SFRFs there are numerous objects with ${\rm SFR_{IR}}$ with values more than ${\rm \sim 100  M_{\odot} \, yr^{-1}}$. 
  
\item In contrast with the UV SFRF the IR SFRF varies with redshift at z =  4 to z = 1. However, overall the derived ${\rm CSFRD_{UV+IR}}$ (using the double-power laws of table \ref{table22}) remains almost constant at this epoch except for some variations at z = 3 and z = 1.5 (z = 4 - 0.106 ${\rm \frac{M_{\odot} \, yr^{-1}}{Mpc^3}}$, z = 3 - 0.123  ${\rm \frac{M_{\odot} \, yr^{-1}}{Mpc^3}}$, z = 2 - 0.107  ${\rm \frac{M_{\odot} \, yr^{-1}}{Mpc^3}}$, z = 1.5 - 0.133  ${\rm \frac{M_{\odot} \, yr^{-1}}{Mpc^3}}$, z = 1 - 0.104  ${\rm \frac{M_{\odot} \, yr^{-1}}{Mpc^3}}$). Like with the case of the ${\rm UV_{corr}}$ CSFRD we see that there is a plateau for the (${\rm CSFRD_{UV+IR}}$) and not a Peak, in agreement with \citet{Gruppioni2020}. At ${\rm z \sim 1}$ to 0.25 according to IR data there is a fast decrement for the high star forming end that causes a decrement to the CSFRD as well (top panel of Fig. \ref{fig:SFRIR1EEE}). From redshift 0.25 to 0 the decrement is much slower,

\item the discrepancies at the high star forming end results in differences between the inferred ${\rm CSFRD_{UV+IR}}$ and ${\rm CSFRD_{UV, corr}}$ evolutions (top right panel of Fig. \ref{fig:SFRIR1EEE}) from 0.3 dex to 0.5 dex. 
  
\end{itemize}

In conclusion, both methods agree on a constant CSFRD at ${\rm z \sim 1-4}$ (plateau) and then a decrement at lower redshifts.  We expect that the constrains given above can be used in comparisons with cosmological simulations like previous measurements \citep{Katsianis2017,Zhaoka2020,Lovell2021}. However, due to the severe differences of the indicators for the high star forming end we can safely state that they cannot both be describing the same galaxy formation and evolution scenario in terms of SFRs. We note that IR SFRs/CSFRDs are not probing {\it necessarily} obscured SFR like it is adopted by numerous studies \citep[e.g.][]{Rowanrobinson2016,Gruppioni2020}. They just have higher\footnote{We have to keep in mind that state-of-the-art SED modelling and radiative transfer simulations actually point that indeed previous measurements/calibrations, especially those of IR, overestimated the derived SFRs \citep{Katsianis2020,Leja2020}.} values than other measurements. We test the legitimacy of the derived ${\rm UV_{corr}}$-SFRFs and UV+IR-SFRFs with respect the independent measurement of the stellar mass density (SMD) from ${\rm z \sim 0-8}$. However, the main goal for the next sections is to determine a parameterization for both the UV+IR and UV$_{corr}$ CSFRDs.

\section{The star formation histories of galaxies and the  cosmic star formation rate density.}
\label{CSFRDintro}

Several studies have focused on parametrizing the SFH of individual galaxies or/and the cosmic SFH with ``physics-free'' models \citep{Abramson2016,Ciesla2017}. The above functions are typically empirically motivated and have been described by the following paremetarizations \citep{Tacchella2018,Chattopadhyay2020}:
\begin{itemize}
  
\item Exponentially declining SFHs \citep{McLure2018} in which star formation jumps from zero to its maximum value $SFR_{0}$ at some time $ T_{o}$. After $T_{o}$ star formation declines exponentially with a time scale $\tau$:
  
\begin{eqnarray}
\label{eq:CSFRMadau0}
{\rm SFR(t) = SFR_{0} \times e^{(-\frac{t-To}{\tau})}}.
\end{eqnarray}

Assuming $T_{o}$ = 0 and $ \lambda = 1/\tau$ produces the standard form of a negative exponentially declining SFH. Short values of $\tau$ correspond to galaxies where most of the stars were formed early on and in a small time, followed by a smooth decrease of the SFR, while high values point to a roughly constant SFR \citep{Ciesla2017}. However, the above parameterization is not appropriate at high redshifts where SFHs of galaxies are expected to rise \citep{reddy2012,Carnall2019}. 

\item Delayed exponentially declining SFHs \citep{Chevallard2019} are considered to be more realistic. After the first generation of stars are formed, some time is required for them to evolve and ultimately end their lives (via supernovae explosions in case of very massive stars). The ejection of material from the first generation supernovae enrich the medium for the second generation of star formation but with delay. The above can be written as:
  
\begin{eqnarray}
\label{eq:CSFRMadau1}
{\rm SFR(t) = SFR_{0} \times (t-T_{o}) \times e^{(-\frac{t-To}{\tau})}.}
\end{eqnarray}

\item the \citet{Yang2013} SFHs of galaxies. In general, the evolution of the SFHs of central galaxies (which drive mostly the CSFRD) is governed by 3 processes: (1) its in situ star formation; (2) the accretion of stars from satellite galaxies; and (3) its passive evolution (mass loss). Since the SFRs of central galaxies depend on halo mass and z in the model, this can be combined with the halo mass function (sum the individual SFRs of galaxies to generate the total) and the CSFRD can be described as:

\begin{eqnarray}
\label{eq:CSFRMadau3}
      {\rm CSFRD = \int_{0}^{\infty} SFR(M_{halo}, z) \, n(M_{halo,}, z) \, d M_{halo}},
\end{eqnarray}  
where $SFR(M_{h},z) = SFR_{pk} \times e^{\frac{log^2(1+z)/(1+z)}{2 \sigma ^2 (z_{pk})}}$ and $\sigma (z_{pk})$ describes the decay of the SFR with respect the peak $SFR_{pk}$. This model has been found to perform well at low redshifts ($z < 2.0$) since it is in good agreement with the observed CSFRD. However, it has some limitations:
  \begin{itemize}
\item The model relies on the adopted stellar mass - halo mass relation to calculate the stellar mass of the central galaxy $M_{\star , 0.1}$ at redshift z = 0.1 \citep{Yang2012} and other functional forms to calculate the peak value ($ SFR_{pk} = \frac{M_{\star , 0.1}}{10^{9.3} h^{-2}}$ ${\rm M_{\odot}/yr}$) of the SFRs within a halo and the redshift at which this peak occurs ($z_{pk} = max[a \times log_{10}(M_{h}-b),0]$). This involves multiple parameters and the functional forms may not represent real galaxies,
\item  At $z>2.0$ the model under-predicts the CSFRD with respect observations.
\end{itemize}

  


\end{itemize}

The above functional forms besides their successes do not typically manage to model the early SFH or/and the position of the peak of SFR is offset with respect recent observations \citep{Ciesla2017}. In addition, as mentioned above they are mostly ``physics-free'' parameterizations which have their roots on empirical motivation \citep{Abramson2016} and sometimes require multiple parameters\footnote{Some other widely used parameterizations with the same limitations are the double power law \citep{Behroozi2013} and the Log-normal \citep{Gladders2013,Abramson2016} SFHs}.

The most widely used way to describe the evolution of the SFR of the Universe as a whole is by providing estimates of the Cosmic Star Formation Rate Density at various redshifts. These can be obtained by integrating Luminosity Functions (UV, IR, H$\alpha$, Radio) or Star Formation Rate Functions \citep{Madau2014,Katsianis2016}. It is a common practice to explore the CSFRD(z) and thus the evolution of the SFR of the Universe by fitting the data with a model that summarize the results. For example, \citet{Madau2014} suggested that the cosmic SFH follows a rising phase, scaling as $ SFR(z) \propto (1 + z)^{-2.9}$ at $ {\rm z = 3 - 8}$, slowing down and peaking at ${\rm z \sim 2}$, followed by a gradual decline to the present day, roughly as $SFR(z) \propto (1 +z)^{2.7}$. The above requires 4 parameters and can be written as:

\begin{eqnarray}
\label{eq:CSFRMadau}
{\rm CSFRD(z) = {\bf 0.015} \frac{(1+z)^{{\bf 2.7}}}{1+[(1+z)/{\bf 2.9}]^{{\bf 5.6}}}} {\rm \frac{M_{\odot} \, yr^{-1}}{Mpc^3}}.
\end{eqnarray}

Equation \ref{eq:CSFRMadau} has been widely used in the literature \citep{Wilkins2019,Maniyar2018,Walter2020} and is represented by the red solid line in the top panel of Fig. \ref{fig:SFRIR1EEE}. An updated version is present at \citet{Madau2017} which requires once again 4 parameters. Similar models have been suggested earlier by \citet{Cole2001} as $CSFRD = \frac{a+b \, z}{1+(\frac{z}{c})^d}$ but 2 questions arise. 1) Is it possible to decrease the number of parameters necessary to broadly fit the data ?  2) Equation \ref{eq:CSFRMadau} represents a ``physics-free' parameterization that is mostly empirical. Is there a physical motivated analytical form/fit that can reproduce the observations and at the same time connect its parameters to the properties of halos/galaxies ?

We remind that in our work in order to obtain the CSFRD(z) we integrate the ${\rm UV_{corr}}$ SFRFs (Schechter forms) and UV+IR SFRFs (double-power forms) of Fig. \ref{fig:SFRIR1} adopting integration limits of SFR = 0.01 - 1000 ${\rm M_{\odot} \, yr^{-1}}$. We summarize our results for the CSFRD at the top right panel of Fig. \ref{fig:SFRIR1EEE} and table \ref{table22}. We adopt a parametric form for the of $ CSFRD(z) = \frac{{\bf C}}{(1+z)^{{\bf D}}} \times e^{{\Huge -\frac{\bf E}{1+z}}} \, {\rm \frac{M_{\odot} \, yr^{-1}}{Mpc^3}}  $ that relies on 3 parameters and the results are for the two indicators as follows:

\begin{eqnarray}
\label{eq:CSFRparaIRt}
{\rm CSFRD}_{\rm UV+IR}(z) = \frac{29.1}{(1+z)^{2.56}} \times e^{{\Huge \frac{-8.07}{1+z}}} \,  {\rm \frac{M_{\odot} \, yr^{-1}}{Mpc^3}},
\end{eqnarray}

\begin{eqnarray}
\label{eq:CSFRparaIR}
      {\rm CSFRD}_{\rm UV,corr}(z) = \frac{79.48}{(1+z)^{3.53}} \times e^{{\Huge \frac{-9.5}{1+z}}} \, {\rm \frac{M_{\odot} \, yr^{-1}}{Mpc^3}}.
\end{eqnarray}

The evolution given by \citet{Madau2014} is in excellent agreement with our ${\rm UV_{corr}}$ CSFRDs for high redshifts (the measurements of the authors rely as well on UV SFRs via UV LFs). However, at lower redshifts \citet{Madau2014} employ a compilation of both IR and UV data. This combination makes the evolution described by the authors to have a sharper peak at ${\rm z \sim 2}$ with respect our ${\rm CSFRD_{UV, corr}}$(z) evolution (black dashed line) and actually slowly converges to IR data (magenta dashed line) at ${\rm z \sim 0}$. As demonstrated in the previous subsection ${\rm UV_{corr}}$ and IR SFRFs are different distributions at the high star forming end so we decide in contrast with \citet{Madau2014} to investigate them separately. Focusing on the ${\rm CSFRD_{UV+IR}}$ of our work we see that we are in good agreement with the results of \citet{Gruppioni2020} which are represented by the red open circles of the top right panel of Fig. \ref{fig:SFRIR1EEE}.  

\section{The ``observed'' cosmic star formation rate density is described by {\bf two} parameters and a function that resembles a Gamma distribution.}
\label{CSFRD}

In order to investigate more physically the evolution of the CSFRD we perform the parameterization/fit with time (Gyrs) at the left panel of Fig. \ref{figEpicG} while the ${\rm CSFRD_{UV+IR}}$ is described as:

\begin{eqnarray}
\label{eq:CSFRparaIRG}
{\rm CSFRD_{UV+IR}}(T) = 0.10 \times \, T^{1.34} \, \times \, e^{-0.43 \, T} \, \, {\rm \frac{M_{\odot} \, yr^{-1} }{Mpc^3}},
\end{eqnarray}

where T is time in Gyrs,  while for the UV corrected data (solid black line of Fig. \ref{figEpicG})  we have:

\begin{eqnarray}
\label{eq:CSFRparauvG}
 {\rm CSFRD}_{\rm UV,corr}(T) = 0.037 \times \, T^{1.83} \, \times \, e^{-0.48 \, T}  \, {\rm \frac{M_{\odot} \, yr^{-1} }{Mpc^3}}.
\end{eqnarray}
The above form for the UV+IR data represents a hybrid of a power law rise (${\rm \sim T^{1.34}}$) and an exponential decline (${\rm \sim e^{-0.43 \, T}}$).  The decreasing amplitude via the exponential decline (${\rm exp^{-C \, T}}$ with characteristic timescale $1/C$ in Gyr) seems that it has origins to the fact that the gas supply to the inner galaxy is depleted with time and reflects gas consumption timescales \citep{Dekel2014,Burkert2017}. Bellow we present how this hybrid form emerges for the CSFRD:

A system that is isolated (a closed box model scenario), has an initial gas mass $ P_{0}$ and no further accretion occurs within it is expected to have a gas density evolution of $P_{gas} = P_{0} \times e^{-b_{\star}  \, t}$, where $b_{\star}= \frac{\epsilon_{SFR}}{\tau_{free \, fall}}$, $\tau_{free \, fall}$ the free fall timescale and $\epsilon_{SFR}$ the SF efficiency per free fall time. This depletion is the result of gas consumption due to star formation which occurs at a rate of $ SFRD =   b_{\star} \times P_{0} \times e^{- b_{\star} \, t}$. All the above reflects the empirical Schmidt law \citep{Schmidt1959,Schaye2008} that relates the SFR and gas via $SFR = b_{\star} \times M_{gas}$. The parameters that govern the decline in eq \ref{eq:CSFRparaIRG} and eq \ref{eq:CSFRparauvG} imply timescales of ${\rm 1/0.43 \, Gyr \sim 2.5 \times 10^9}$ yr for the case of ${\rm CSFRD_{UV+IR}}$ and ${\rm 1/0.48 \, Gyr \sim 2 \times 10^9}$ yr for the case of ${\rm CSFRD_{UV, corr}}$, respectively. These values are typically found in observations of {\it individual} galaxies \citep[][$ \tau = 1-3 \times 10^9 $]{kennicutt1998,Bigiel2011,Davis2015}  and semi-analytic models \citep[][$\tau = 2 \times 10^9 $ yr]{Shamshiri2015} so it is a good start that our observations of the CSFRD find a similar value. As noted above (section \ref{CSFRDintro}) solely the exponential declining scheme is not successful at reproducing the rising SFHs of individual high redshift galaxies. Similarly to the individual SFHs, the CSFRD is anticipated to increase with time at ${\rm z \ge 3}$. This era has been labelled as the ``gas accretion'' epoch and although there have been different ways to parametrize this behavior, the most straightforward way is by a simple power law ($SFH(T) = A \times T^B$) following \citet{Papovich2011} and \citet{Dutton2010}. The value of $ B$ has been found  ${\rm \sim 1.5}$ by \citet[semi-analytic models]{Shamshiri2015}, ${\rm \sim 1.7}$ by \citet[][high redshift ($z \sim 6-3$) observations]{Papovich2011} and ${\rm \sim 3.5}$ by \citet[][model]{Behroozi2013}. We find $ B = 1.34$ for the case of ${\rm CSFRD_{UV+IR}}$ (eq. \ref{eq:CSFRparaIRG}) and $B = 1.83$ for the case of  ${\rm CSFRD_{UV, Corr}}$ (eq. \ref{eq:CSFRparauvG}). We have to note that the hybrid form of the above processes (power law rise + exponential declining), i.e. $CSFRD(T) = A \times T^B \times e^{ \, b \, T}$, has not been commonly used to parameterize the {\it observed} CSFRD for the whole history of the Universe ($z \sim 0-9$) despite its success in describing our data. We can see the parameterization/model as a more general case of the delayed exponential declining SFH ($SFR_{0} \times t \times e^{(-\frac{t}{\tau})}$) where this time the rising/delaying part $SFR_{0} \times t$ has an extra parameter B. This parameter sets how quickly the SFR rise at the accretion era {\it and} since the gas is finite sets the limit for the exponential part to take over. For $B = 1$ we have the simple case of delayed exponential decaying SFH. We will gain more insight on the role of these parameters at the following paragraphs.

Interestingly if we fit the UV+IR data with two parameters we see that equation \ref{eq:CSFRparaIRG} is almost identical (see Fig. \ref{figEpicG}) to the following form:

\begin{figure*}
\centering
\includegraphics[scale=0.45]{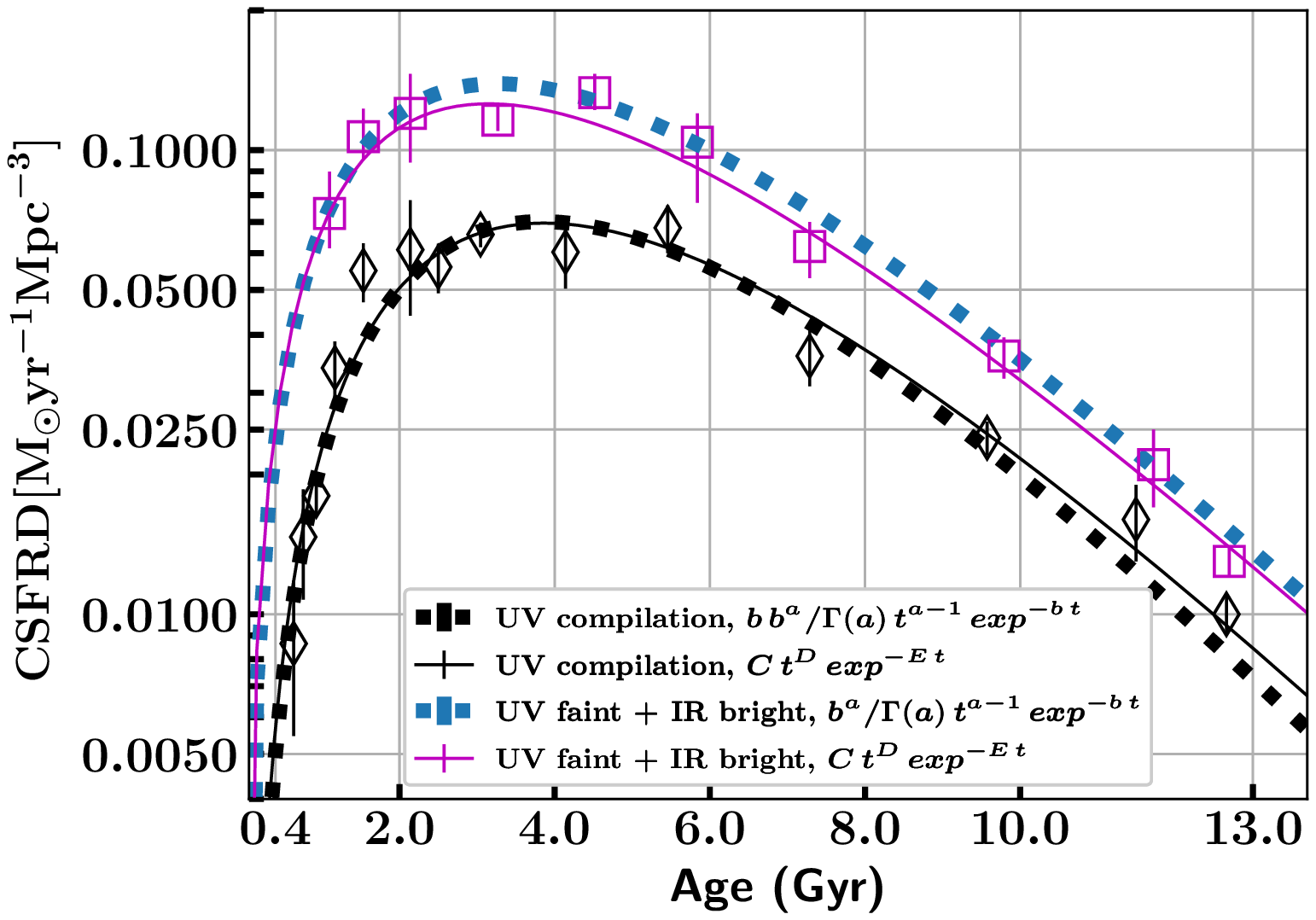}
\includegraphics[scale=0.45]{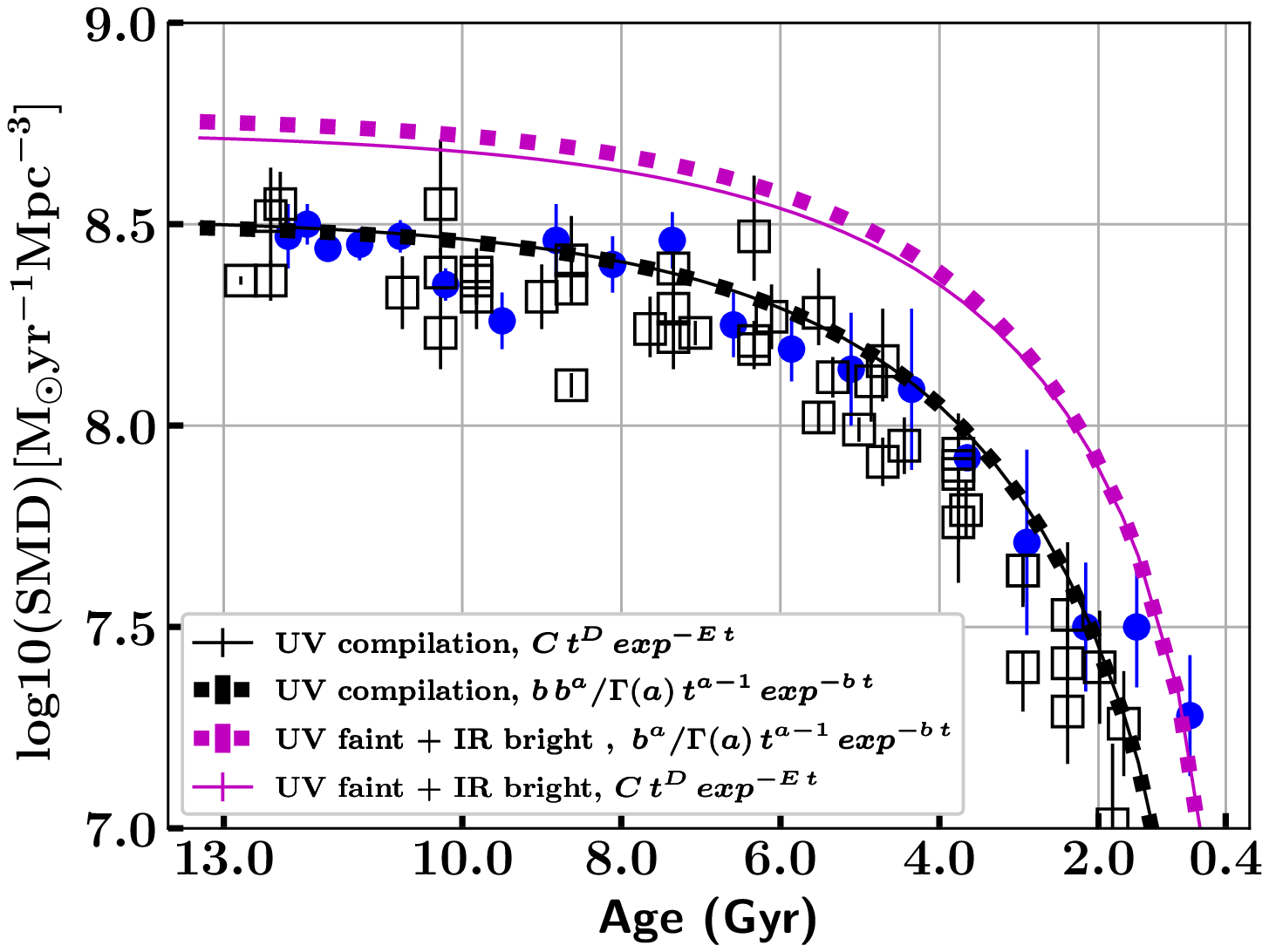}
\vspace{-0.50cm}
\caption{Left panel: The evolution of the Cosmic Star Formation Rate Density with time (in Gyr). Parameterizations (employing 3 or 2 free parameters) for both UV and IR data are present.  Right panel: The evolution of the stellar mass density, derived by integration over time of the parameterizations of the left panel alongside a compilation of observations \citep{Bielby2012,Ibert2013,Muzzin2013,Madau2014,Driver2018}. We see that the UV CSFRD evolution is in agreement with the stellar mass density evolution, while IR can deviate from 0.3 dex to 0.5 dex.}
\label{figEpicG}
\end{figure*}

\begin{eqnarray}
\label{eq:GAMAIR}
{\rm CSFRD}_{\rm UV+IR} (T) \approx  \frac{{\bf 0.44^{2.44}}}{\Gamma({\bf 2.44)}} \times T^{{\bf 2.44}-1} \times e^{-{\bf 0.44} \, T } \, \, {\rm \frac{M_{\odot} \, yr^{-1} }{Mpc^3}},
\end{eqnarray}

where T is time in Gyrs. The above is represented by the  dotted  magenta line of Fig. \ref{figEpicG} and deviates from the actual fit with 3 parameters only by ${\rm 0}$ to ${\rm 0.05}$ dex.
This is a function that resembles closely the formula of a Gamma distribution which has the following behavior:

\begin{eqnarray}
\label{eq:GAMAIRgga}
{\rm Gamma(a,b) = \frac{b^{a}}{\Gamma(a)} \times x^{a-1} \times e^{-bx}}, 
\end{eqnarray}

where $a$ is the shape parameter and $b$ is the scale parameter. We shall see that the ${\rm UV_{corr}}$ CSFRD follows a similar behavior. The scale $ b$ parameter describes the rate $(1/b)$ that the process occurs and has the effect of stretching or compressing the range of the distribution. The parameter $ a$ determines its skewness. The gamma distribution is commonly used to model stochastic processes and an example is in reliability analysis where partial failures (which are gamma distributed) have to occur before the item completely fails. The parameter $a$ describes the number of events necessary to generate the complete failure. A natural question that arises is how the dependence of the CSFRD on time is dictated by this elegant mathematical formula. Why the normalization of equation \ref{eq:CSFRparaIRG} is in the end just a product of $a$ and $b$ ? We reproduce the above by a simple analysis inspired by a bath tub / equilibrium model that follows the evolution of the gas/SF in section \ref{Justification} and produce a physical motivated fit for the cosmic star formation rate density with the parameters $a$ and $b$ being determined by halo/galaxy properties. 

\subsection{A comparison between the ${\rm UV_{corr}}$/UV+IR CSFRDs and stellar mass density evolution.}

Before studying the emergence of the above delicate form we can start seeing its physical meaning, while we have to note that a huge advantage in modeling processes through gamma functions is that any related mathematical calculations becomes straightforward and can be calculated analytically. It is a reason why both the Gamma function and the Gamma distribution find numerous applications from biology to engineering \citep{Maghsoodloo2014,Vazquez2020,McCombs2020}. For example, integrating the above and multiplying it by a factor of $(1-R)$, where $R = 0.41$ is the return fraction for a \citet{chabrier03,Madau2014} IMF (i.e. the mass fraction of each generation of stars that is put instantly back into the interstellar/intergalactic medium), gives us the build up of the stellar mass in the Universe\footnote{Besides that it is a common practice in the literature to adopt a constant return fraction $R$ with respect time \citep{MunozMateos2007,Madau2014,Grazian2015,Walter2020} the mass fraction is not returned instantly. Instead, the return fraction is sensitive to the age of the stellar population and is a function of time. For example, \citet{Yu2016} employed the Flexible Stellar Population Synthesis (FSPS) code \citep{Conroy2010} to derive an evolving return fraction $R(t)$. The authors demonstrated that the SMD derived from the CSFRD using the evolving $R(t)$ instead of the constant $R$ is larger with increasing redshift by up to a factor of 1.2 at $z = 8$.} :

\begin{eqnarray}
\label{eq:GAMAIRintegr}
{\rm SMD_{UV+IR}(T)} = (1-R) \times \frac{b^{a}}{\Gamma(a)} \times \int_{0}^{t} \, T^{a-1} \times e^{-bT} \, dt = (1-R) \times \frac{\gamma(a,b \, T)}{\Gamma(a)} \times {\rm 10^9  \, {\rm \frac{M_{\odot}}{Mpc^3}}},
\end{eqnarray}

where t is time in yrs and $ \gamma(a,b \, T)$ is the lower incomplete gamma function. The above is represented by the magenta dashed line of the right panel of Fig. \ref{figEpicG}. So UV+IR SFRs point to a build up of cosmic stellar mass that occurs following an incomplete gamma function that is {\it normalized} by the total Gamma function (i.e. the regularized gamma function). The ``ultimate fate'' of the stellar mass of the Universe at $T = \infty$ is according to equation \ref{eq:GAMAIRintegr}, $ {\rm SMD_{UV+IR}(\infty)}$ $\sim 0.59 \times  {\rm 10^9  {\rm \, M_{\odot}} \, Mpc^{-3}}$ (since at $T = \infty$, $\frac{\gamma(a,b \, T)}{\Gamma(a)} = 1 $) while the stellar mass density grows following {\it gamma distributed increments}.  Mathematically it seems like partial stochastic events occur before the final event and total collapse of the system. The black open squares of the right panel of Fig. \ref{figEpicG} represent the observational studies of \citet{Bielby2012}, \citet{Ibert2013}, \citet{Muzzin2013}, \citet{Madau2014} and \citet[][blue filled circles]{Driver2018} that are obtained through the SED fitting technique. The solid magenta line represents the evolution of the Stellar mass density integrating the 3 parameter function (eq. \ref{eq:CSFRparaIRG}) which closely follows the result from the 2 parameter function. We note that the UV+IR SFRs have a discrepancy with respect stellar masses by ${\rm \sim 0.3}$ dex at ${\rm z \sim 0}$ while this increases to ${\rm \sim 0.5}$ dex at higher redshifts (z ${\rm \sim }$ 2). What we see is the emergence once again of the long-standing factor of 2-3 disagreement between CSFRD and Stellar mass density \citep{Leja2015,Davidzon2018}. However, as mentioned above, \citet{Katsianis2020} using EAGLE combined with the radiative transfer code SKIRT \citep{Baes2020,Trcka2020} has demonstrated that numerous UV+IR calibrations that are used to determine galaxy SFRs \citep{Heinis2014,Whitaker2014,Tomczak2016} typically overestimate the results at high redshifts. In addition, \citet{Leja2019b} inferred via Prospector SFRs which are lower by 0.1-1 dex with respect ${\rm SFR_{UV+IR}}$ measurements \citep{Whitaker2014}. The authors suggest that this is due to the inclusion of additional physics (e.g. light from old stars). In this work we show that the observed UV+IR SFRs and observed stellar masses are compared and are found to be in disagreement. Stellar mass is an accumulating property that is highly dependent on the past (i.e. high redshifts). Overestimating the SFRs at early epochs results in an overestimation of the stellar mass density for all the history of the Universe {\it including ${\rm z \sim 0}$ which is a widely accepted constrain that has been commonly employed by the community}\footnote{The GSMF (which integration produces the SMD) is usually the key calibration for most galaxy formation models \citep{Leauthaud2011,Schaye2015,Wright2017,Pillepich2018} since it is considered robust.}. We note that our result is independent on the parameterization of our model (which just facilitates the procedure of calculating the stellar mass density). But what about the ${\rm UV_{corr}}$ SFRFs and CSFRD?

Following the same steps with the ${\rm CSFRD_{UV+IR}}$ this time we explore the evolution of the UV dust corrected SFRs (black line of right panel of Fig. \ref{fig:SFRIR1EEE}) which can be written in the form :

\begin{eqnarray}
\label{eq:GAMAUVV}
{\rm CSFRD}_{\rm UV,corr} \approx  0.50 \times \frac{0.50^{2.9}}{\Gamma(2.9)} \times T^{2.9-1} \times e^{-0.50\, T} \, {\rm \frac{M_{\odot} \, yr^{-1} }{Mpc^3}}, 
\end{eqnarray}

which is shown at the left panel of Fig. \ref{figEpicG}, that once again requires only two parameters and integrating it gives us the evolution of the cosmic stellar mass density according to the UV + dust corrected SFRs as follows:

\begin{eqnarray}
\label{eq:GAMAUVintegr}
{\rm SMD_{UV,corr}(T)} = (1-R) \times b \times \frac{b^{a}}{\Gamma(a)} \times \int_{0}^{t} \, T^{a-1} \times e^{-b \, T} \, dt = (1-R) \times b \times \frac{\gamma(a,b \, T)}{\Gamma(a)} {\rm \, M_{\odot}\times 10^9 \, {\rm \frac{M_{\odot}}{Mpc^3}}}.
\end{eqnarray}

The above is represented by the black dotted line of the right panel of Fig. \ref{figEpicG}. The black solid line of the same figure describes the stellar mass density derived from the integration of the relation \ref{eq:CSFRparauvG} that involves 3 parameters instead of 2. The ${\rm UV_{corr}}$ results this time (in contrast with the UV+IR SFRs) are consistent with the observed stellar mass densities of Fig. \ref{figEpicG}. We note the following:

\begin{itemize}

  
\item The observed UV+IR SFRs imply cosmic star formation rate densities that if integrated with time result in a stellar mass density evolution inconsistent with other measurements that are derived through SED fitting. These measurements involve the whole history of the Universe, including the GSMF at $z \sim 0$ which represents constraints for numerous models and a ``cornerstone'' for extragalactic astrophysics. The tension increases at high redshifts even by up to 0.5 dex re-creating the long-standing factor of 2-3 disagreement. However, the case is different for the CSFRD derived from ${\rm SFR_{UV_{corr}}}$. We suggest that the success of the  indicator at high redshifts, also confirmed by simulations/radiative transfer, where dust attenuation effects are less severe leads to a good agreement between the observed ${\rm CSFRD}$ derived from the dust corrected UV data and SMD. The above agreement represents a suggestion for the long-standing problem of the tension between observed CSFRDs and SMD which seems to be actually the result of the uncertainties of SFR indicators and {\it not} a problem related to the theory of galaxy formation. 
  
\end{itemize}
  
\section{A simple physical model that reproduce the observed CSFRD {\bf and} links its parameters to the physics of halos and galaxies.}
\label{Justification}

There has been considerable effort in cosmological simulations \citep{TescariKaW2013,Crain2015,Pillepich2018,Lagos2018,dave2019} {\it and} theory \citep{Yang2013,Dekel2014,PengMaolino2014,Sharma2020} to create galaxies that resemble the ones we observe. The cosmic star formation rate density has always been one of the key observables that constrain the above. Bellow we demonstrate how the ``Gamma CSFRD'' present in this work is obtained using a simple model. 

We start with the SFR within galaxies. We assume that the total baryonic mass is conserved and is separated between the available gas for star formation within the galaxies ($M_{SF, T}$), stellar mass ($M_{\star, t}$), outflows ($M_{out, t}$) and inflows ($M_{in, t}$). Star-forming galaxies in simulations are usually seen to lie near the equilibrium condition \citep{Dave2012,PengMaolino2014} where star formation is sustained by the inflowing gas $M_{in, t}$, thus we assume the same for our simple model. We divide by volume (labelled as Vol) in order to work with densities and all the above can be written as :

\begin{eqnarray}
\label{eq:PGgast}
{\rm \frac{P_{in, t}}{dt} = \frac{P_{out, t}}{dt} + \frac{P_{\star, t}}{dt}},
\end{eqnarray}

where $ P_{in, t} = \frac{M_{in, t}}{Vol}$, $P_{out, t} = \frac{M_{out, t}}{Vol}$ and $P_{\star, t} = \frac{M_{\star, t}}{Vol}$. Star-forming galaxies fluctuate around this relation but are generally driven back to it on short timescales. We assume that the outflows are dependent on the SFR, thus the star formation rate density labelled as SFRD is related to $P_{out, t}$ as $\frac{ \, P_{out, t}}{dt} = n \times SFRD$, where n is the mass loading factor \citep{springel2003,barai13,PuchweinSpri12,Katsianis2016,Lopez2020,Tejos2021} and $\frac{ \, P_{\star, t}}{dt} = (1-R) \times SFRD$. Thus, the SFRD is related to the inflows $P_{in, t}$ as:

\begin{eqnarray}
\label{eq:PGgast00}
{\rm SFRD = \frac{dP_{in, t}}{dt}\frac{1}{1+n-R}.}
\end{eqnarray}

The inflowing baryonic gas is assumed to scale with the DM halo growth as $\frac{dP_{in, t}}{dt} = f_{gal} \times f_{b} \times \frac{dP_{Halo}, t}{dt}$, where $f_{b}$ is the cosmic baryon fraction and $f_{gal}$ is the fraction of the incoming baryons that they are able to penetrate and reach the galaxy and be used for star formation \citep{PengMaolino2014}. $f_{gal}$ is equivalent to the accretion efficiency defined by \citet{Bouche2010}, the preventive feedback parameter of \citet{Dave2012} and the penetration parameter p presented by \citet{Dekel2014}. Thus, equation \ref{eq:PGgast00} can be written as:

\begin{eqnarray}
\label{eq:PGgast01}
{\rm SFRD = \frac{f_{gal} \times f_{b}}{1+n-R} \times \frac{dP_{Halo, t}}{dt}.}
\end{eqnarray}
The above relation connects the SFRD of a galaxy to the growth of the halo.

We now focus on the cosmic SFRD labeled as the CSFRD. According to eq. \ref{eq:CSFRparaIRG} we initially adopt a dependence of the CSFRD on 3 parameters. As described in the previous sections the parameterization that describes the cosmic star formation rate density as $CSFRD(T) = A \, T^{B} \, \times \, e^{-b_{\star} \, T}  \, \frac{M_{\odot} \, yr^{-1} }{Mpc^3}$ is the natural result of the gas accretion era which occurs at early epochs and is governed by a power law (the parameter B) with an exponential decline that takes over at lower redshifts (via the parameter $b_{\star}$).  We shall see the physical meaning of this form as we explore its parameters. We convert the above into $CSFRD = A \, t^{B} \, \times \, e^{-b_{\star}\, t} \,  =  A \, t^{a-1} \, \times \, e^{-b_{\star} \, t}  \, \frac{M_{\odot} \, yr^{-1} }{Mpc^3}$, where $B = a-1$ and remind that  $b_{\star} = \frac{\epsilon_{SFR}}{\tau_{free \, fall}}$ is the star formation depletion timescale, $\tau_{free \, fall}$ the free fall timescale and $\epsilon_{SFR}$ the SF efficiency per free fall time.

We can investigate the relation of the constant A (the normalization of the $CSFRD = A \, t^{B} \, \times \, e^{-b \, t}$)  with parameters that have a physical origin by extending equation \ref{eq:PGgast01} for the cosmic SFRD (CSFRD) and the total dark matter accretion rate within halos ($P_{C, Halo, t}$). We assume that SFR = 0 at t = 0 and we obtain the following:

\begin{eqnarray}
\label{eq:PGgast}
{\rm A \times \int_{0}^{t} \, T^{a-1} \times \, e^{-b \, T} dt= f_{gal} \times f_{b} \times P_{C, Halo, t} \times \frac{1}{1+n-R}.}
\end{eqnarray}

Integrating up to $t =  \infty $ we have :

\begin{eqnarray}
\label{eq:PGgast011}
{\rm {\bf A = \frac{b^{a}}{\Gamma(a)} \, \frac{ f_{gal} \times f_{b} \times P_{C, Halo, \infty}}{10^9 \times (1+n-R)}.}}
\end{eqnarray}

The above just reflects the endgame for the whole baryonic gas of the Universe that is mostly accreted to Halos and trapped in stars and stellar remnants (except the $1-f_{gal}$ fraction). We assume $f_{b} = 0.157$, all the eligible ($f_{gal}$) dark matter at ${\rm \infty}$ has been accreted in dark matter halos so $ P_{C, Halo, \infty} = \Omega_{Matter} \times P_{crit}$ where $P_{crit}$ is the critical density of the Universe ($\frac{3 \, H_{o}^2}{8 \pi G}$) and $\Omega_{Matter} = 0.315 $  is the matter density parameter. Thus, substituting $P_{C, Halo, \infty} = 0.315 \times \frac{3 \, H_{o}^2}{8 \pi G} = 0.315 \times 1.245 \times 10^{11} \, \frac{M_{\odot}}{Mpc^3}$, mass loading factor $n = 2$\footnote{In our simple model we adopt constant galactic winds, with a fixed wind mass loading factor n = 2. We note that for individual halos a more appropriate assumption would be to associate outflows with the circular velocity i.e. $n =  2 \times (\frac{450 \, Km/s}{2 \times V_{circular}})^{2}$ \citep{springel2003,barai13,PuchweinSpri12,Katsianis2016}.} and the return fraction R = 0.41 the A parameter can be written as:

\begin{eqnarray}
\label{eq:PGgast1}
{\rm A = \frac{b^{a}}{\Gamma(a)} \, \frac{ f_{gal} \times f_{b} \times \Omega_{M} \times P_{crit} }{(1+n-R)}  = \frac{b^{a}}{\Gamma(a)} \, \frac{ f_{gal} \times 0.0494 \times 1.245 \times 10^{11}}{10^9 \times (1+2-0.41)}  =  \frac{b^{a}}{\Gamma(a)} \times f_{gal} \times 2.37},
\end{eqnarray}

where $f_{gal}$ is considered to have a value of ${\rm \sim 0.5}$ \citep{PengMaolino2014,Dekel2014}. \citet{Dave2012} suggest that the preventative parameter $f_{gal}$ is Halo dependent while \citet{Katsianis2017} has demonstrated using EAGLE that most of the  CSFRD bellow redshift 5 occurs in Halos between ${\rm 10^{11} M_{\odot} }$ to ${\rm 10^{12} M_{\odot}}$ so we can focus our analysis at this halo range. At this mass regime values of ${\rm 0.2}$ to ${\rm 0.5}$ are expected for the $f_{gal}$ parameter according to \citet{Dave2012}. A realistic value of $f_{gal} = 0.45 \sim 0.5$ that is mostly used in the literature would result to a value of $A \sim \frac{b^{a}}{\Gamma(a)} $ and be in perfect agreement with the relation \ref{eq:GAMAIR}, the UV+IR CSFRD and its Gamma distribution evolution. A lower value of $f_{gal} \sim 0.25$ would result to the relation implied by the CSFRD tha relies on dust corrected UV data. $f_{gal}$ has a maximum of 1 so it is anticipated that the parameter A has a maximum of $\frac{b^{a}}{\Gamma(a)} \times 2.37$. We note that both values $f_{gal} = 0.45$ and $f_{gal} = 0.25$ are within the constrains from cosmological simulations. These values explain the emerging forms for the CSFRD of equations \ref{eq:GAMAIR} and \ref{eq:GAMAUVV} and result to an evolution for the star formation rate density that can be described mostly by two parameters despite the fact that numerous physical processes are involved. We note that the 3 parameters equation is physically motivated, while the 2 parameters fit occurs due to the fact that the parameters of galaxy formation physics are the way they are i.e. the $f_{gal}$, $f_{b}$, $\Omega_{Matter}$ and $P_{crit}$ parameters (which all 4 determine the accreting baryonic mass) balance out the outflow mass loading factor n and the return fraction R in order for the $\frac{ f_{gal} \times f_{b} \times \Omega_{M} \times P_{crit} }{(1+n-R)}$ ratio to be 1 (UV+IR data, equation \ref{eq:GAMAIR}) or 0.5 (dust corrected UV data, equation \ref{eq:GAMAUVV}). More work is needed to distinguish which case represents better the Universe, if this cancellation of the parameters happens by ``chance'' and multiple scenarios could be involved. 

Since we have determined the parameters of the fit and linked them to physical properties (A is given by equation \ref{eq:PGgast011},  $ b = \frac{\epsilon_{SFR}}{\tau_{free \, fall}}$ where $\tau_{free \, fall}$ the free fall timescale and $\epsilon_{SFR}$ the SF efficiency per free fall time) we can explore now the physical meaning of the form. The CSFRD of the Universe can simply be seen as a gas consumption rate ($\frac{d \, P_{\star}}{dt} = - \frac{d \, P_{Gas}}{dt}$) that occurs in $\Gamma(a,b)$ distributed steps up to the point that there is no eligible gas that is able to penetrate/infall at $\infty$ i.e:

\begin{eqnarray}
\label{eq:PGgastconsumption0}
{\rm {\bf CSFRD =  \frac{ f_{gal} \times f_{b} \times \Omega_{M} \times P_{crit}}{(1+n-R)} \,  \times \frac{b_{\star}^{a}}{\Gamma(a)} \, \times \, T^{a-1} \times e^{-b_{\star} \, T }  \, {\rm \frac{M_{\odot} \, Gyr^{-1} }{Mpc^3}}}}.
\end{eqnarray}

We change the form of the above in order to see it in the shape of a differencial equation ($\frac{d \, P_{\star}}{dt} = - \frac{d \, P_{Gas}}{dt}$) as :

\begin{eqnarray}
\label{eq:PGgastconsumption1}
{\rm  CSFRD =  - \frac{ f_{gal} \times f_{b} \times \Omega_{M} \times P_{crit}}{(1+n-R)} \,  \times \frac{d}{dT}\left(\sum_{x=0}^{a-1} \frac{(b_{\star} \, T)^x \, e^{-b_{\star} \, T}}{x!} \right)  \, {\rm \frac{M_{\odot} \, Gyr^{-1} }{Mpc^3}}}.
\end{eqnarray}

For $a = 1$ we have an exponential declining SFH described as $CSFRD(T) =  \frac{ f_{gal} \times f_{b} \times \Omega_{M} \times P_{crit}}{(1+n-R)} \times  b \, e^{-b \, T} \, \frac{M_{\odot} \, Gyr^{-1}}{Mpc^3}$. This reflects a scenario that gas does not need to be accreted and no furher delay occurs caused by the fact that the consumption is separated between different generations. Gas just collapses, starts being consumed and stars start emerging. For $a$ ${\rm >}$ 1 the SFH evolves following equation \ref{eq:PGgastconsumption1}. This is an evolution where star formation/gas consumption occurs at different steps and at early stages is governed by a rising component driven by gas accretion. The above processes are determined by the parameter $a$. The physical interpretation derived in this subsection describes successfully  the CSFRD while it connects the parameters of the fit to the properties of galaxies and halos.

\section{Conclusions}
\label{con}

In this paper we employ state-of-the-art UV and IR datasets to investigate the star formation rate function (SFRF) and cosmic star formation rate density (CSFRD) at $z \sim  0-9$. In agreement with previous studies we find that IR and ${\rm UV_{corr}}$ derived SFRFs contradict each other on some key issues, mostly due to their differences at the high star forming end. However, both tracers have been offering very useful insights about galaxy evolution for decades and interesting results are driven from our analysis that connects observations to a simple theoretical model and an interesting parameterization:

\begin{itemize}

\item both tracers agree on a constant CSFRD evolution at $z \sim 1-4$ and point to a plateau instead of a peak at $z \sim 2$,
  
\item the observed CSFRD can be described by 3 parameters ($C \times T^{a-1}\times e^{-b \, T}$) and follows a hybrid of a power law rise and an exponential decline. Individual galaxies have been seen to follow broadly this parameterization via what is commonly labelled the delayed star formation history (SFH) model. Exploring further the CSFRD we see that actually it is mostly governed by {\it two} parameters and a {\it function that resembles a Gamma distribution} ($ \frac{b^{a}}{\Gamma(a)} \times T^{a-1} \times e^{-b \, T}$ / $b \times \frac{b^{a}}{\Gamma(a)} \times T^{a-1} \times e^{-b \, T}$).

\item We support that the parameterization of the CSFRD with 2-3 parameters is interesting because it reflects physics related to galaxy formation. In contrast to previous efforts our framework connects the parameters to physical properties (SFR depletion times, cosmic baryonic gas density, penetration parameter from the halo to the galaxy e.t.c.). The build up of stellar mass in our model occurs in ${\rm \Gamma(a,b)}$ distributed steps and is the result of gas consumption up to the limit that there is no available gas at $t = \infty$,

\item the final value of the cosmic stellar mass density is ${\rm \sim 0.5 \times 10^9 \, \frac{M_{\odot}}{Mpc^3}}$ at $t = \infty$. We are approaching this value so we can infer that most of the stars that were ``supposed'' to be born from the available baryonic gas have already done so,                      
  
\item The observed UV+IR SFRFs imply cosmic star formation rate densities that if integrated with time result in a stellar mass density evolution from $z \sim 0-9$ inconsistent with measurements that are obtained through SED fitting. The tension increases at high redshifts even by up to 0.5 dex. This could be seen as the long-standing problem of the tension between observed CSFRDs and stellar mass densities. Does it reflect the fact that high redshift SFRs derived from UV+IR calibrations are overestimated ? The above is supported by radiative transfer simulations and SED fitting techniques \citep{Katsianis2020}). On the other hand the ${\rm SFR_{UV,corr}}$ indicator produces SFRFs and a CSFRD evolution which is actually consistent with the stellar mass density evolution and the problem of the long-standing tension can be solved.
  
\end{itemize}

We are obligated to stress once again that any observational techniques (including the ones employed in this work) rely on assumptions and calibrations that can be incomplete so we caution any other author to not treat our (and theirs) ``observed'' SFRs as the ``true'' SFRs. There is considerable work to be done both in terms of data and modelling until we can disentangle if UV/SED/H$\alpha$ or IR/Radio/CO are more robust at high-z. We stress that any models, simulations and parameterizations (including ours) are by construction affected by the limitations of the observational techniques.

\section*{Acknowledgments}

We  thank  the  anonymous  referee  for  their suggestions  and  comments  that  improved  the  quality  of  our Paper. A.K has been supported by the Tsung-Dao Lee Institute Fellowship and Shanghai Jiao Tong University. X.Y. is supported by the national science foundation of China (grant Nos. 11833005,11890692,11621303) and Shanghai Natural Science Foundation,grant No.15ZR1446700. We also thank the support of the Key Laboratory for Particle Physics, Astrophysics and Cosmology, Ministry of Education. XZZ acknowledges the supports from the National Key Research and Development Program of China (2017YFA0402703),the National Science Foundation of China (11773076, 12073078),and the Chinese Academy of Sciences (CAS) through a China-Chile Joint Research Fund (CCJRF no: 1809) administered by the CAS South America Centre for Astronomy (CASSACA)

\bibliographystyle{apj}		
\bibliography{Katsianis_ApJ8}


\label{lastpage}
\end{document}